\documentclass[12pt]{article}
\usepackage{amsfonts}
\usepackage{amssymb}
\usepackage{amsmath}
\usepackage{latexsym}
\usepackage{epsfig}
\usepackage{dsfont,eucal,mathrsfs}

\newif{\ifcomentarios}
\comentariosfalse

\newtheorem{theorem}{Theorem}
\newtheorem{lemma}[theorem]{Lemma}

\newtheorem{proposition}[theorem]{Proposition}

\newtheorem{remark}[theorem]{Remark}

\renewcommand{\mathbf}{\boldsymbol}
\renewcommand{\mathcal}{\mathscr}

\begin{document}

\author{\textbf{Domingos H. U. Marchetti}\thanks{%
Partially supported by CNPq and FAPESP. E-mail: \textit{\ marchett@if.usp.br}%
} \ \ and \ \ \textbf{William R. P. Conti}\thanks{%
Supported by FAPESP under grant $\#98/10745-1$. E-mail: \textit{\
william@if.usp.br}. \ \ \ \ \ \ } \\
%EndAName
Instituto de F\'{\i}sica\\
Universidade de S\~{a}o Paulo \\
Caixa Postal 66318\\
05315 S\~{a}o Paulo, SP, Brasil \and \textbf{Leonardo F. Guidi}\thanks{%
E-mail: \textit{\ guidi@mat.ufrgs.br}} \\
%EndAName
Instituto de Matem\'{a}tica \\
Universidade Federal do Rio Grande do Sul \\
Av. Bento Gon\c{c}alves, 9500 - Pr\'{e}dio 43-111 \\
91509 Porto Alegre, RS, Brasil}
\title{Hierarchical Spherical Model from a Geometric \\
Point of View}
\date{}
\maketitle

\begin{abstract}
A continuous version of the hierarchical spherical model at dimension $d=4$
is investigated. Two limit distribution of the block spin variable $%
X^{\gamma }$, normalized with exponents $\gamma =d+2$ and $\gamma =d$ at and
above the critical temperature, are established. These results are proven by
solving certain evolution equations corresponding to the renormalization
group (RG) transformation of the $O(N)$ hierarchical spin model of block
size $L^{d}$ in the limit $L\downarrow 1$ and $N\rightarrow \infty $.
Starting far away from the stationary Gaussian fixed point the trajectories
of these dynamical system pass through two different regimes with
distinguishable crossover behavior. An interpretation of this trajectories
is given by the geometric theory of functions which describe precisely the
motion of the Lee--Yang zeroes. The large--$N$ limit of RG transformation
with $L^{d}$ fixed equal to $2$, at the criticality, has recently been
investigated in both weak and strong (coupling) regimes by Watanabe \cite{W}%
. Although our analysis deals only with $N=\infty $ case, it complements
various aspects of that work.
\end{abstract}

\section{Introduction and Statement of Results\label{I}}

\setcounter{equation}{0} \setcounter{theorem}{0}

We continue the investigation starting in \cite{CM}. In the present work we
give a geometric interpretation to certain trajectories of a first order
partial differential equation related to the renormalization group
transformation (RGT) of a $d$--dimensional hierarchical spherical model.

\medskip

\noindent \textbf{Motivation.} The hierarchical $O(N)$ spin model, with $%
L^{d}=2$ sites per block, has been recently studied by renormalization group
in both weak and strong regimes by Watanabe \cite{W}. Starting from the
uniform \textquotedblleft a priori\textquotedblright\ measure supported in
the $N$--dimensional sphere of radius $\sqrt{N}$, the critical trajectory of
the RGT has shown to converge to the Gaussian fixed point for sufficiently
large $N$. To control such trajectory, which starts far away from the fixed
point, the exactly solved $O(\infty )$ trajectory has been used together
with two key ingredients: reflection positivity and the Lee--Yang property
of\ single--site\ \textquotedblleft a priori\textquotedblright\ measures.
The former ingredient gives uniform convergence of $O(N)$ trajectories to $%
O(\infty )$ trajectories. The latter property has been previously employed
by Kozitsky \cite{K} to establish two central limit theorems. Watanabe's
analysis, based in his joint work with Hara and Hattori \cite{HHW} on the
critical trajectory for the hierarchical Ising model ($N=1$), in
contradistinction to Kozitsky's, and most of the previous studies of this
model, does not restrict the space of \textquotedblleft a
priori\textquotedblright\ measures to a neighborhood of the Gaussian fixed
point and is able to deal with the borderline $d=4$ case.

Although the analysis of the RGT with $L^{d}\geq 2$ fixed is expected to be
simplified considerably in the $L\downarrow 1$ limit (see e.g. \cite{F}),
none of the above mentioned results can be carried to the limit as the two
key ingredients do not hold if $L^{d}$ is not an integer. In order to
establish, in the local potential approximation ($L\downarrow 1$), a weak
convergence of the hierarchical $O(N)$ Heisenberg equilibrium measure to the
corresponding spherical equilibrium measure as $N\rightarrow \infty $ an
entirely new method of analysis has to be developed from scratch.

In the present investigation we establish central limit theorems for the
four--dimensional hierarchical spherical ($N=\infty $) model at and above
the critical temperature. Our results are achieved in the local potential
approximation that reduces the renormalization group equation to a nonlinear
first order partial differential equation. A geometric function
interpretation of the $O(\infty )$ trajectory is thus given with the help of
an explicit solution obtained by the method of characteristics. It follows
from our analysis that the Lee--Yang zeroes reach a limit distribution as
the Gaussian fixed point approaches but their support moves away to infinity.

\medskip

\noindent \textbf{The model.} The hierarchical Heisenberg model on a finite
box $\Lambda _{K}=\left\{ 0,1,\ldots ,L^{K}-1\right\} ^{d}\subset \mathbb{Z}%
^{d}$ of size $n=L^{dK}$ is given by the $O(N)$ invariant equilibrium measure%
\begin{equation}
d\nu _{n}^{(N)}(\mathbf{x})=\frac{1}{Z_{n}^{(N)}}\exp \left\{ \frac{1}{2}%
\left( \mathbf{x},A\mathbf{x}\right) _{\Omega _{n}}\right\}
\prod_{j=1}^{n}d\sigma _{0}^{(N)}\left( x_{j}\right)  \label{equ}
\end{equation}%
where $\mathbf{x}=\left( x_{1},\ldots ,x_{n}\right) $ denotes an element of
the configuration space $\Omega _{n}=\mathbb{R}^{N}\times \cdots \times 
\mathbb{R}^{N}$; $A=J\otimes I$ the tensor product of the hierarchical
coupling matrix $J$ (whose quadratic form 
\begin{eqnarray}
\left( s,-Js\right) _{\Lambda _{K}} &=&-\left( L-1\right)
~\sum_{k=1}^{K}L^{-2k}\sum_{r\in \Lambda _{K-k}}\left( B^{k}s\right)
_{r}^{2}~,  \notag \\
\left( Bs\right) _{i} &=&\frac{1}{L^{d/2}}\sum_{j\in \left\{ 0,\ldots
,L-1\right\} ^{d}}s_{Li+j}~,  \label{block}
\end{eqnarray}%
coincides with Dyson's hierarchical energy \cite{D} when there are $L^{d}=2$
sites per block)\footnote{%
The factor $L-1$ is chosen so that the hierarchical Laplacean converges, as $%
L\downarrow 1$, to a continuum hierarchical Laplacean (see \cite{F,CM}).}
with the $N\times N$ identity matrix $I$; $\sigma _{0}\left( x\right) $ the
\textquotedblleft a priori\textquotedblright\ uniform measure on the $N$%
--dimensional sphere $\left\vert x\right\vert ^{2}=\beta N$ of radius $\sqrt{%
\beta N}$ with $\beta $ the inverse temperature.

\medskip

\noindent \textbf{Recursion relations.} The invariance of $J$ under block
transformation (\ref{block}) allows to establish a recursion relation: 
\begin{equation}
\sigma _{k}^{(N)}(x)=\frac{1}{C_{k}}e^{c_{\gamma }(L-1)\left\vert
x\right\vert ^{2}/2}\underset{L^{d}-\mathrm{times}}{\underbrace{\sigma
_{k-1}^{(N)}\ast \cdots \ast \sigma _{k-1}^{(N)}}}(L^{\gamma /2}x)~~
\label{sigmak}
\end{equation}%
on the space of single--site \textquotedblleft a priori\textquotedblright\
measures in $\mathbb{R}^{N}$ with initial data $\sigma _{0}^{(N)}(x)$. Here, 
$\ast $ denotes the convolution product 
\begin{equation*}
\rho \ast \eta (x)=\int_{\mathbb{R}^{N}}\rho (x-x^{\prime })\,d\eta
(x^{\prime })~,
\end{equation*}%
$C_{k}$ is chosen so that $\sigma _{k}^{(N)}$ is a probability measure and 
\begin{equation}
c_{\gamma }=\left\{ 
\begin{array}{lll}
1 & \mathrm{if} & \gamma =d+2 \\ 
L^{-2k} & \mathrm{if} & \gamma =d%
\end{array}%
\right. ~.  \label{c}
\end{equation}%
The \textquotedblleft a priori\textquotedblright\ measure\ $\sigma
_{k}^{(N)} $ at the step $k$ is defined by integrating (\ref{equ}) over $%
\Omega _{n}$ with the value of $k$--th block spin fixed: 
\begin{equation*}
\int \delta \left( \left( B^{k}\otimes I\right) \mathbf{y}-\mathbf{x}\right)
~d\nu _{n}^{(N)}(\mathbf{y})=\frac{1}{Z_{L^{d(K-k)}}^{(N)}}\exp \left\{ 
\frac{1}{2}\left( \mathbf{x},A\mathbf{x}\right) _{\Omega
_{L^{d(K-k)}}}\right\} \prod_{j=1}^{L^{d(K-k)}}d\sigma _{k}^{(N)}\left(
x_{j}\right) ~
\end{equation*}%
is a marginal measure on $\Omega _{L^{d(K-k)}}$ that preserves the form (\ref%
{equ}).

In terms of their characteristic functions 
\begin{equation}
\phi _{k}^{(N)}(z)=\int \exp \left( ix\cdot z\right) ~d\sigma _{k}^{(N)}(x)~,
\label{cf}
\end{equation}%
equation (\ref{sigmak}) reads%
\begin{equation}
\phi _{k}^{(N)}(z)=\frac{1}{N_{k}}\exp \left( \frac{-L+1}{2}c_{\gamma
}\Delta \right) \left( \phi _{k-1}^{(N)}(L^{-\gamma /2}z)\right) ^{L^{d}}
\label{rr}
\end{equation}%
for $k\geq 1$ with%
\begin{equation}
\phi _{0}^{(N)}(z)=\frac{\Gamma \left( N/2\right) }{\left( \sqrt{\beta N}%
\left\vert z\right\vert /2\right) ^{N/2-1}}J_{N/2-1}\left( \sqrt{\beta N}%
\left\vert z\right\vert \right) :=\varphi _{0}^{(N)}(\left\vert z\right\vert
)~.  \label{ic}
\end{equation}%
Here, $\exp \left( t\Delta \right) $ is the semi--group generated by the $N$%
--dimensional Laplacean operator $\Delta =\partial ^{2}/\partial
z_{1}^{2}+\cdots +\partial ^{2}/\partial z_{N}^{2}$, $N_{k}$ is chosen so
that $\phi _{k}(0)=1$ holds for all $k=1,\ldots ,K$ and $J_{\alpha }(x)$ is
the Bessel function of order $\alpha $ (see eq. ($20$) in Chapter $V\!I\!I$
of \cite{CH} for an appropriate integral representation). Note that $\phi
_{k}^{(N)}(z)=\varphi _{k}^{(N)}(r)$ depends only on $r=\left\vert
z\right\vert =\sqrt{z\cdot z}$.

\medskip

\noindent \textbf{Thermodynamical functions.} The macroscopic behavior of
the model is described by the limit distribution of the block variable 
\begin{equation}
X_{n,N}^{\gamma }=\frac{1}{\sqrt{n^{\gamma /d}}}\sum_{j=1}^{n}x_{j}~,
\label{sum}
\end{equation}%
where $\gamma $ is chosen in order the limit law to be attained. The
characteristic function associated with the block variable $X_{n,N}^{\gamma
} $ with $\gamma =d+2$ is given by 
\begin{eqnarray*}
\Phi _{n}^{(N)}\left( z\right) &=&\int \exp \left( iL^{-K(d+2)/2}\left(
\sum_{j=1}^{n}x_{j}\right) \cdot z\right) d\nu _{n}^{(N)}(\mathbf{x}) \\
&=&\int \exp \left( ix\cdot z\right) ~d\sigma _{K}^{(N)}(x)=\varphi
_{K}^{(N)}(\left\vert z\right\vert )~.
\end{eqnarray*}%
As $n$ goes to infinite, $X_{n,N}^{\gamma }$ converges in distribution to $%
X_{N}^{\gamma }$ if $\varphi _{K}^{(N)}(r)$ converges at every point $r\geq
0 $ to a function $\varphi ^{(N)}(r)$ that is continuous at $r=0$, by
continuity theorem (see e.g. \cite{D}). The convergence of $\nu
^{(N)}=\lim_{n\rightarrow \infty }\nu _{n}^{(N)}$ to the equilibrium measure 
$\nu $ of the spherical model is more subtle and we analogously employ: $%
X_{N}^{\gamma }$ is said to converges to $X^{\gamma }$ in distribution if 
\begin{equation*}
\lim_{N\rightarrow \infty }\left( \varphi ^{(N)}(\sqrt{N}r)\right)
^{1/N}=\varphi (r)
\end{equation*}%
exist for every point $r\geq 0$, is continuous at $r=0$ and coincides with
the corresponding characteristic function of the spherical model. The
reescaling is seen to be necessary already at the initial function $\varphi
_{0}^{(N)}(r)$ (see Proposition \ref{visclim}).

The statements about convergence are independently of which order both
limits $n\rightarrow \infty $ and $N\rightarrow \infty $ are taken. This has
been shown in \cite{CM} adapting a method employed by Kac and Thompson \cite%
{KT} for\ the hierarchical equilibrium measure (\ref{equ}) with $\gamma =d$, 
$L^{d}\geq 2$ an integer and $\beta $ different from the critical inverse
temperature $\beta _{c}=\beta _{c}(d,L)$ of the hierarchical spherical
model. In \cite{CM}, $X^{d}$ is shown (see Theorem $2.3$ and Remarks $4.2$)
to be Gaussian with mean zero and variance $1/\mu $ where $\mu =\mu (\beta )$
is implicitly defined by%
\begin{equation}
\beta =\int \frac{1}{\lambda -\mu }d\varrho (\lambda )  \label{implicity}
\end{equation}%
where $\varrho (\lambda )$ is the density of eigenvalues (counted
multiplicities) of the hierarchical Laplacean $-\Delta _{H}=\dfrac{L-1}{%
L^{2}-1}I-J$. Note $J$ is not invariant under translation by a vector in $%
\mathbb{Z}^{d}$, property that is required for coupling matrices in \cite{KT}%
. Some statements about hierarchical spherical model hold also in the limit
as $L\downarrow 1$, in which case (\ref{implicity}) reads (see Section $3$
of \cite{CM})\footnote{%
By monotonicity, there exist a unique solution $\mu =\mu (\beta )<0$ defined
for $0<\beta <4$.}%
\begin{equation}
1-\frac{\beta }{4}=-2\mu \ln \left( 1-\frac{1}{2\mu }\right) ~
\label{beta-mu}
\end{equation}%
for $d=4$. Central limit theorems are established in the present work
directly from the $L\downarrow 1$ limit.

\medskip

\noindent \textbf{Local Potential Approximation.} Let 
\begin{equation}
U(t,z)=-\ln \phi _{k}^{(N)}(z)  \label{U}
\end{equation}%
be defined for $t=k\ln L$. As $k\rightarrow \infty $ together with $%
L\downarrow 1$ so that $k\ln L$ is kept fixed at a positive real number $t$,
(\ref{c}) converges to%
\begin{equation*}
c_{\gamma }(t)=\left\{ 
\begin{array}{lll}
1 & \mathrm{if} & \gamma =d+2 \\ 
e^{-2t} & \mathrm{if} & \gamma =d%
\end{array}%
\right. ~
\end{equation*}%
and we have 
\begin{eqnarray*}
U_{t} &=&\lim_{L\downarrow 1}\frac{U(t,z)-U(t-\ln L,z)}{\ln L} \\
&=&\lim_{k\rightarrow \infty }\frac{k}{t}\left\{ -\ln \left[ \frac{1}{N_{k}}%
\exp \left\{ -\frac{t}{2k}c_{\gamma }\Delta \right\} \left( \phi
_{k-1}^{(N)}\left( e^{-\gamma t/2k}z\right) \right) ^{e^{td/k}}\right] +\ln
\phi _{k-1}^{(N)}(z)\right\} ~.
\end{eqnarray*}%
Consequently, (\ref{U}) satisfies the initial value problem 
\begin{equation}
U_{t}=-\frac{c_{\gamma }}{2}\left( \Delta U-\left\vert U_{z}\right\vert
^{2}\right) +dU-\frac{\gamma }{2}z\cdot U_{z}+\frac{c_{\gamma }}{2}\Delta
U(t,0)\,  \label{V}
\end{equation}%
with 
\begin{equation}
U(0,z)=-\ln \phi _{0}^{(N)}(z)~.~  \label{V0}
\end{equation}%
The last term in the right hand side ensures that $U(t,0)=0$ for all $t\geq
0 $. Note that this property is satisfied by the initial condition because
of the normalization $\int \sigma _{0}^{(N)}(dx)=\phi _{0}^{(N)}(0)=1$.

We shall prove two limit theorems (Theorems \ref{gauss} and \ref{normal})
summarized as 
\begin{equation*}
\lim_{t\rightarrow \infty }\lim_{N\rightarrow \infty }\frac{1}{N}U(t,\sqrt{N}%
z)=\left\{ 
\begin{array}{lll}
\left\vert z\right\vert ^{2} & \mathrm{if} & \beta =\beta _{c} \\ 
-\left\vert z\right\vert ^{2}/2\mu & \mathrm{if} & \beta <\beta _{c}%
\end{array}%
\right. ~
\end{equation*}%
uniformly in compact subsets of $\zeta \in \mathbb{C}$ with $\Re e\left(
\zeta \right) =-\left\vert z\right\vert ^{2}$. The first, when the sum (\ref%
{sum}) is normalized with abnormal exponent $\gamma /d=1+2/d$, holds at the
critical point 
\begin{equation}  \label{betac}
\beta =\beta _{c}(d)=\dfrac{2d}{d-2} \; ,
\end{equation}
$d\geq 4$. The second, for normal exponent $\gamma /d=1$, holds for any $%
\beta <\beta _{c}(d)$ and $d>2$. In both cases only the borderline $d=4$
will be considered for brevity.

\medskip

\noindent \textbf{Conformal mapping.} Although continuity at $\left\vert
z\right\vert =0$ suffices for these limit theorems, the \textquotedblleft
characteristic function\textquotedblright\ \ $\lim_{N\rightarrow \infty
}\exp \left( \dfrac{-1}{N}U(t,\sqrt{N}z)\right) $ is shown to be an analytic
function that converges, as $t\rightarrow \infty $, to an entire function.
In addition, thanks to an explicit solution of the initial value problem (%
\ref{V}) and (\ref{V0}) at $N=\infty $, the whole trajectory can be
described by the geometric function theory.

The initial value (\ref{V0}) is a function of $\left\vert z\right\vert ^{2}$
and equation (\ref{V}) preserves this property. So, we define 
\begin{equation}
u(t,x)=\lim_{N\rightarrow \infty }\frac{1}{N}U(t,\sqrt{N}z)  \label{uU}
\end{equation}%
for $x=-\left\vert z\right\vert ^{2}$ and let, for each $t\geq 0$, the
partial derivative $u_{x}(t,\zeta )$ of $u$ be extended as an analytic
function of $\zeta =x+iy$ with $y>0$. We prove in Theorem \ref{conformal}
that $u_{x}\left( t,\zeta \right) $,$~t\geq 0$, map the upper half--plane $%
\mathbb{H}$ conformally into a decreasing family of open convex sets 
\begin{equation*}
u_{x}\left( t,\mathbb{H}\right) =\Omega _{t}\subset \Omega _{0}=u_{x}\left(
0,\mathbb{H}\right)
\end{equation*}%
contained in $\mathbb{H}$, and there is a one--to--one and onto relation
between this family and the trajectory $\mathcal{O}$ at the critical inverse
temperature $\beta _{c}(4)=4$ converging to the Gaussian fixed point.
Analogous theorem holds for the trajectory corresponding to normal
fluctuations.

The boundary of $\Omega _{t}$ is the union of a segment $I_{\alpha
}:=[-\alpha ,0]$ extending from a point $-\alpha =-\alpha (t)<0$ up to the
origin over the real line and a convex\ curve $q=h(t,p)$, $p\in I_{\alpha }$%
, with $h(t,-\alpha )=h(t,0)=0$. $h(t,I_{\alpha })=\left\{ h(t,p),p\in
I_{\alpha }~\right\} $ encodes all informations about $\mathcal{O}$ since it
corresponds to the image of a branching cut of $u_{x}(t,\zeta )$. The
principal branch of $u_{x}(t,\zeta )$ belongs to the Pick class of
analytical function and admits to be represented as 
\begin{equation}
u_{x}\left( t,\zeta \right) =-1+\int_{-\infty }^{\infty }\left( \frac{1}{%
\lambda -\zeta }-\frac{1}{\lambda -1/2}\right) ~d\mu (t,\lambda )~
\label{vmu}
\end{equation}%
where $d\mu =\rho ~d\lambda $ is absolutely continuous (with respect to
Lebesgue) Borel measure. Although (\ref{vmu}) is not a canonical
representation, 
\begin{equation*}
\rho \left( t,\lambda \right) =\frac{1}{\pi }\lim_{\eta \downarrow 0}\Im
\left( u_{x}\left( t,\lambda +i\eta \right) \right)
\end{equation*}%
holds as well. Denoting by $\Sigma (t)=\left( -\infty ,-d(t)\right) $ the
support of $\mu $ in (\ref{vmu}), we have 
\begin{eqnarray*}
-\alpha (t) &=&u_{x}\left( t,-d(t)\right) \\
h(t,I_{\alpha }) &=&\Im \left( u_{x}\left( t,\Sigma (t)+i0\right) \right)
\end{eqnarray*}

The support $\Sigma (t)$ of $\mu (t,\lambda )$ determines the location of
the Lee--Yang zeroes as it can be seen by representing $\varphi
_{k}^{(N)}(r) $ into a infinite canonical product (for $\varphi _{0}^{(N)}(r)
$, see proof of Proposition \ref{visclim}). By (\ref{U}) and (\ref{uU}),
these zeroes are poles of $u_{x}(t,\zeta )$ that become dense over\ the
semi--line $\Sigma (t) $ as $N\rightarrow \infty $. As $t$ goes to $\infty $%
, $\alpha (t)\rightarrow 3/2$, $d(t)\rightarrow \infty $ leading $\Sigma (t)$
to an empty set $\emptyset $ as all Lee-Yang singularities are expelled to
infinite. As a consequence, $u_{x}\left( t,\zeta \right) \rightarrow -1$
uniformly in each compact set of $\mathbb{C}$.

The motion of the Lee--Yang zeroes can be attained from the moments of their
distribution \cite{N}. The moments satisfy an infinite system of ordinary
first--order differential equations which is reduced in \cite{HHW,W} to a
finite system by Lee--Yang inequalities. The presence of one--dimension
unstable manifold makes this system very sensitive to truncation and no
simplification occurs in the limit $N\rightarrow \infty $. This has to be
contrasted with the simple geometric analysis in Section \ref{GFT} from
which the dynamics of Lee--Yang zeroes can be described globally.

\medskip

\noindent \textbf{Outline.} In Sections \ref{CLTEE} and \ref{NF} we prove
Theorems \ref{gauss} and \ref{normal}, which are Gaussian limit laws for the
spherical model on the local potential approximation. Section \ref{GFT}
presents an interpretation of explicit solution of the associate nonlinear
first order partial differential equation according to the geometric
function theory. A conclusion with final remarks is given in Section \ref{CR}%
.

\section{Central Limit Theorem \label{CLTEE}}

\setcounter{equation}{0} \setcounter{theorem}{0}

\noindent \textbf{The radial equation.} The initial value (\ref{V0}) is a
function of $\left\vert z\right\vert ^{2}=r^{2}$ and the spherical symmetry
is preserved by the evolution equation (\ref{V}). So, it suffices to take
into account the radial component of $z\cdot \partial /\partial z$ and $%
\Delta $, respectively given by $r\partial /\partial r$ and 
\begin{equation*}
\frac{1}{r^{N-1}}\frac{\partial }{\partial r}\left( r^{N-1}\frac{\partial }{%
\partial r}\right) =\frac{\partial ^{2}}{\partial r^{2}}+(N-1)\frac{1}{r}~%
\frac{\partial }{\partial r}~.
\end{equation*}

Defining%
\begin{equation}
u^{(N)}\left( t,x\right) =\frac{1}{N}U(t,\sqrt{N}z)  \label{uV}
\end{equation}%
for $x=-\left\vert z\right\vert ^{2}$, the initial value problem (\ref{V})
and (\ref{V0}) for $\gamma =d+2$ reads 
\begin{equation}
u_{t}^{(N)}=\frac{2}{N}xu_{xx}^{(N)}+u_{x}^{(N)}-2x\left( u_{x}^{(N)}\right)
^{2}-\gamma xu_{x}^{(N)}+du^{(N)}-u_{x}^{(N)}(t,0)  \label{u}
\end{equation}%
with $u^{(N)}\left( 0,x\right) =U(0,\sqrt{N}z)/N$. As $N\rightarrow \infty $
, the initial function converges to a limit:

\begin{proposition}
\label{visclim} 
\begin{equation}
\lim_{N\rightarrow \infty }u^{(N)}\left( 0,x\right) =\int_{0}^{x}\frac{%
-\beta }{1+\sqrt{1+4\beta x^{\prime }}}dx^{\prime }~\equiv u_{0}(x)
\label{u0}
\end{equation}%
and the convergence is uniform in any compact set of the slit plane $\mathbb{%
C}\backslash (-\infty ,-1/4\beta ]$.
\end{proposition}

Proposition \ref{visclim} is proven in Section \ref{GFT}. Watanabe
established (\ref{u0}) writing $u_{0}(x)$ as a continued fraction of Gauss
(see Lemma $4.1.$of \cite{Wa}). Additional properties are obtained by taking
into account that $u_{0}^{\prime }$ is an analytic function of the Pick
class $P_{I(\beta )}$ which is able to be continued across the interval $%
I(\beta )=\left( -1/4\beta ,\infty \right) $.

\medskip

\noindent \textbf{Viscosity limit equation}\footnote{$1/N$ plays the role of
viscosity since it is in front of the Laplacean as in the hydrodynamic
equation of incompressible fluid. Viscosity solution (or limit) also refers
to a method for obtaining \textquotedblleft weak
solutions\textquotedblright\ of semilinear first order partial differential
equations (see e.g. \cite{E}).}. Taking $N\rightarrow \infty $ in (\ref{u})
we are led to a first\ order partial differential equation for (\ref{uU}) 
\begin{equation}
u_{t}=u_{x}-2xu_{x}^{2}-\gamma xu_{x}+du-u_{x}(t,0)  \label{vl}
\end{equation}%
which can be solved by the method of characteristics. To avoid dealing with
a nonlinear equation we apply the Legendre transformation to (\ref{vl}). Let%
\begin{equation}
w(t,p)=\max_{x\geq 0}\left( xp-u(t,x)\right) =\bar{x}p-u(t,\bar{x})
\label{wu}
\end{equation}%
be the Legendre transform of $u$ with respect to $x$ where $\bar{x}=\bar{x}%
(t,p)$ is attained at the value $x$ for which 
\begin{equation}
p=u_{x}(t,x)  \label{p}
\end{equation}%
has a solution for every $t\geq 0$ and $p$ in a certain domain depending on $%
t$.

Assuming $w(t,p)$ continuously differentiable and uniformly convex function
of $p$ such that $\lim_{p\rightarrow \infty }w(t,p)/\left\vert p\right\vert
=\infty $ holds for all $t\geq 0$, the original function $u(t,x)$ can be
recovered by inverse Legendre transformation%
\begin{equation}
u(t,x)=\max_{p\in \mathbb{R}}\left( xp-w(t,p)\right) =x\bar{p}-w(t,\bar{p})
\label{uw}
\end{equation}%
where $\bar{p}=\bar{p}(t,x)$ solves $x=w_{p}\left( t,p\right) $ for $p$.
Note that, by differentiating (\ref{wu}) with respect to $t$ and $p$
together with (\ref{p}), we have%
\begin{eqnarray}
w_{t} &=&-u_{t}  \notag \\
w_{p} &=&\bar{x}+\left( p-u_{x}(t,\bar{x})\right) \bar{x}_{p}=\bar{x}~.
\label{wp}
\end{eqnarray}%
Hence, $w_{p}$ solves equation (\ref{p}) for $x$. We are going to show that $%
w_{p}(t,p)$ is a monotone increasing function of $p$ for every $t\geq 0$
therefore, $w(t,p)$ is convex and a well defined Legendre transform of $u$
which, by (\ref{uw}), is also uniformly convex. It follows by duality of the
Legendre transformation that%
\begin{equation}
\bar{p}(t,x)=u_{x}(t,x)  \label{pbar}
\end{equation}%
which, in view of the presence of $u_{x}(t,0)$ in (\ref{vl}), gives%
\begin{equation}
u(t,x)=\int_{0}^{x}\bar{p}(t,x^{\prime })~dx^{\prime }~.  \label{up}
\end{equation}

Using $\gamma =d+2$ together with (\ref{wu}) and (\ref{wp}), equation (\ref%
{vl}) becomes 
\begin{equation*}
w_{t}=-p+2p\left( 1+p\right) w_{p}+dw+\bar{p}_{0}
\end{equation*}%
where $\bar{p}_{0}=\bar{p}_{0}(t)$ is implicitly defined by the equation $%
0=w_{p}(t,p)$. Writing $v=w_{p}=\bar{x}$ we arrive, by differentiating both
sides of the above equation with respect to $p$, at the following initial
value problem 
\begin{equation}
v_{t}-2p\left( 1+p\right) v_{p}=-1+\left( \gamma +4p\right) v  \label{v}
\end{equation}%
with%
\begin{equation}
v(0,p)=\frac{1}{2p}+\frac{\beta }{4p^{2}}\equiv v_{0}(p)~.  \label{v0}
\end{equation}%
Note that $v(0,p)=\bar{x}(0,p)$ is the value $x$ that solves (\ref{p}) at $%
t=0$: 
\begin{equation}
p=u_{x}(0,x)=u_{0}^{\prime }(x)=\frac{-\beta }{1+\sqrt{1+4\beta x}}
\label{pux}
\end{equation}%
by (\ref{wp}) and (\ref{u0}).

\medskip

\noindent \textbf{Main result.} Our main result of this section is as follows

\begin{theorem}
\label{gauss}Equations (\ref{v}) and (\ref{v0}) with $d=4$ ($\gamma =6$) are
solved by%
\begin{equation}
v(t,p)=\frac{1}{2p}+\frac{1}{p^{2}}-\frac{4-\beta }{4p^{2}}e^{2t}-\frac{1+p}{%
p^{3}}\ln \left( 1+p-pe^{2t}\right) ~.  \label{vtp}
\end{equation}%
At $\beta =\beta _{c}=4$, there is a unique solution $\bar{p}=\bar{p}(t,x)$
of%
\begin{equation}
v(t,p)=x  \label{vx}
\end{equation}%
holomorphic in a neighborhood of origin, that converges, as $t\rightarrow
\infty $, to $-1$ in every compact set of $\mathbb{C}$. Together with
equations (\ref{up}) and (\ref{uV}), this implies convergence to the
Gaussian equilibrium solution of (\ref{V}):%
\begin{equation*}
\lim_{t\rightarrow \infty }\lim_{N\rightarrow \infty }\frac{1}{N}U(t,\sqrt{N}%
z)=\left\vert z\right\vert ^{2}~
\end{equation*}%
uniformly in compacts.
\end{theorem}

\begin{remark}
The use of Legendre transform in the renormalization group transformation
for the $O(N)$ Heisenberg model in the large--$N$ limit goes back to
Shang--Keng Ma's work (see \cite{Ma} and references therein). It is also
reminiscent of the method of Laplace (see eqs. (3.1.13)-(3.1.16) of \cite%
{KKPS}). In ref. \cite{W}, Watanabe solved the discrete flow equation (\ref%
{rr}) with $L^{d}=2$ in the $N\rightarrow \infty $ limit and partial
differential equation is employed only for the heat semigroup part in (\ref%
{rr}). Theorem \ref{gauss} extends Watanabe's result to the flow equation (%
\ref{V}) at the $L\downarrow 1$ limit.
\end{remark}

\begin{remark}
Theorem \ref{gauss} treats the border case $d=4$ but holds for any $d\geq 4$%
. The proof of the theorem can also be adapted to deal with the convergence
to nontrivial equilibrium solutions of (\ref{V}) at $\beta =\beta _{c}(d)$,
given by (\ref{betac}), for $2<d<4$.{}
\end{remark}

\noindent \textbf{Proof.} Theorem \ref{gauss} will be proven by solving (\ref%
{v}) along the characteristics $p(t)=p(t;p_{0})$ (see e.g. \cite{E}).
Writting $V(t)=v(t,p(t))$, equation (\ref{v}) is reduced to a pair of
ordinary differential equations%
\begin{eqnarray}
\dot{p} &=&-2p\left( 1+p\right)  \notag \\
\dot{V} &=&-1+\left( 6+4p\right) V  \label{pt}
\end{eqnarray}%
satisfying initial conditions $p(0)=p_{0}$ and%
\begin{equation}
V(0)=V_{0}=v_{0}(p_{0})~.  \label{pt0}
\end{equation}%
Integrating the first equation of (\ref{pt})%
\begin{equation*}
\int_{p_{0}}^{p}\frac{dp^{\prime }}{p^{\prime }(1+p^{\prime })}%
=\int_{p_{0}}^{p}\left( \frac{1}{p^{\prime }}-\frac{1}{1+p^{\prime }}\right)
dp^{\prime }=-2\int_{0}^{t}dt^{\prime }
\end{equation*}%
gives%
\begin{equation}
p(t)=\frac{p_{0}e^{-2t}}{1+p_{0}-p_{0}e^{-2t}}~.  \label{pp0}
\end{equation}

The second equation of (\ref{pt}) is a nonhomogeneous linear equation. The
homogeneous equation $\dot{V}=\left( 6+4p\right) V$ can be integrated:%
\begin{eqnarray*}
V(t) &=&V_{0}\exp \left( 6t+4\int_{0}^{t}p(s)~ds\right) \\
&=&V_{0}e^{6t}\left( 1+p_{0}-p_{0}e^{-2t}\right) ^{2}~.
\end{eqnarray*}%
Using the variation of constants formula (see Theorem $3.1$ of \cite{CL}),
the solution to the second equation of (\ref{pt}) is given by%
\begin{equation}
V(t)=e^{6t}\left( 1+p_{0}-p_{0}e^{-2t}\right) ^{2}\left( V_{0}-J_{0}\right)
\label{Vt}
\end{equation}%
with%
\begin{equation*}
J_{0}=\int_{0}^{t}\frac{e^{-6s}~ds}{\left( 1+p_{0}-p_{0}e^{-2s}\right) ^{2}}%
~,
\end{equation*}%
by changing variable $\zeta =e^{-2s}$, given by%
\begin{equation*}
J_{0}=\frac{1}{2p_{0}^{3}}\left[ (1+p_{0})^{2}\frac{1}{1+p_{0}-p_{0}\zeta }%
+2(1+p_{0})\ln \left( 1+p_{0}-p_{0}\zeta \right) +p_{0}\zeta \right] _{\exp
\left( -2t\right) }^{1}.
\end{equation*}%
After some manipulations together with (\ref{pt0}) and (\ref{v0}), this
gives 
\begin{equation}
V_{0}-J_{0}=\frac{\beta -6}{4p_{0}^{2}}-\frac{1}{2p_{0}^{3}}+\frac{%
(1+p_{0})^{2}}{2p_{0}^{3}\left( 1+p_{0}-p_{0}e^{-2t}\right) }+\frac{e^{-2t}}{%
2p_{0}^{2}}+\frac{1+p_{0}}{p_{0}^{3}}\ln \left( 1+p_{0}-p_{0}e^{-2t}\right) .
\label{V0J0}
\end{equation}%
Equation (\ref{vtp}) follows by plugging this result into (\ref{Vt}) with $%
p_{0}$ as a function of $t$ and $p$: 
\begin{equation*}
p_{0}(t,p)=\frac{pe^{2t}}{1+p-pe^{2t}}~
\end{equation*}%
obtained by solving (\ref{pp0}) for $p_{0}$.

\medskip

\noindent \textbf{The inverse function theorem.} We now solve equation (\ref%
{vx}) for $p$ at the critical point $\beta =\beta _{c}(4)=4$. By (\ref{vtp}%
), it can be written as 
\begin{equation}
xp^{2}-\frac{p}{2}-1=-\frac{1+p}{p}\ln \left( 1+p-pe^{2t}\right) \equiv
g(t,p)~.  \label{g}
\end{equation}%
The first of two ingredients we need is

\begin{lemma}
\label{monotone}For $p<\left( e^{2t}-1\right) ^{-1}$, $g$ is a monotone
increasing function of $p$ diverging to $-\infty $ logarithmically as $%
p\rightarrow -\infty $ and satisfying $g(t,-1)=0$ and $g(t,0)=\left(
e^{2t}-1\right) $.
\end{lemma}

\noindent \textbf{Proof of lemma.} Clearly, $g$ is well defined function of $%
p$ for $1+p-pe^{2t}=1-p\left( e^{2at}-1\right) >0$ with logarithmic
divergence at $p=-\infty $. We have, by an explicit computation, 
\begin{equation*}
g_{p}(t,p)=\frac{e^{2t}-1}{1-p\left( e^{2t}-1\right) }+\frac{1}{p^{2}}%
f\left( p\left( e^{2t}-1\right) \right)
\end{equation*}%
where%
\begin{equation}
f(w)=\ln \left( 1-w\right) +\frac{1}{1-w}-1\equiv h(w)-1~.  \label{fh}
\end{equation}%
If $f(w)\geq 0$ for all $w<1$ then $g_{p}(t,p)>0$ in the domain $p<\left(
e^{2t}-1\right) ^{-1}$ and the monotonicity statement is proven. In fact, $%
h(0)=1$ and 
\begin{equation*}
h^{\prime }(w)=\frac{w}{\left( 1+w\right) ^{2}}
\end{equation*}%
implies that $w=0$ is the absolute minimum of $h$ proving an equivalent
statement: $h(w)>1$ for $w<1$ different from $0$.

$\hfill \Box $

For $x\leq 0$, the quadratic polynomial 
\begin{equation*}
Q(x,p):=xp^{2}-\frac{p}{2}-1
\end{equation*}%
in the left hand side of (\ref{g}) is bounded from above by a linear
function: 
\begin{equation*}
Q(x,p)\leq Q(0,p)=-\frac{p}{2}-1,
\end{equation*}%
and attains its maximum value $\dfrac{-1}{16x}-1$ at $p_{\mathrm{\max }}=%
\dfrac{1}{4x}$. Since $p_{\max }\rightarrow 0$ as $x\rightarrow -\infty $,
there is a value $x_{\mathrm{\max }}=x_{\mathrm{\max }}(t)$ such that no
real solutions of (\ref{vx}) exist for $x<x_{\mathrm{\max }}$. On the other
hand, as the graph of $g(t,p)$ intercepts the graph of $Q(x,p)$ in two
points (one point) for any $0>x>x_{\mathrm{\max }}(t)$ ($x\geq 0$) and $%
t\geq 0$, there exist at least one real solution of (\ref{vx}) for $x\geq
x_{ \mathrm{\max }}$ (see Figure ). We shall discard the solution associated
with the second point of interception since it diverges at $x=0$.

\begin{figure}[th]
\begin{center}
\epsfig{file=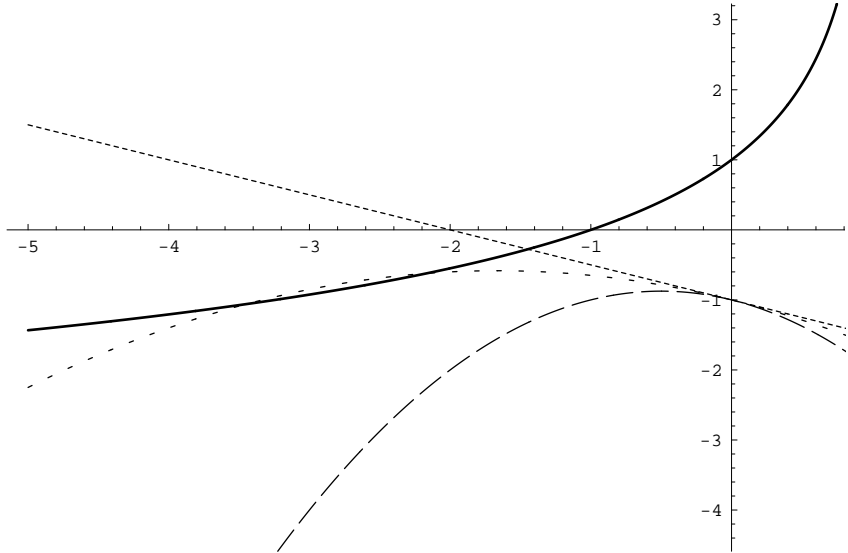, scale=0.7
}
\end{center}
\caption{Graphic of intersections of $g (t,p)$ (solid line) with $Q (x,p)$
for $x=0$ (dotted line) and $x\not = 0$ (dashed lines) }
\label{scpp}
\end{figure}

Now, let $t\geq 0$ and let $x$ and $p$ be real parts of numbers in $\mathbb{%
C }$: $z=x+iy$ and $\eta =p+iq$. Although the solution $\eta =\eta (t,z)$ of 
$z=v(t,\eta )$ is a multivalued function of $z$, only one branch, denoted by 
$\bar{\eta}(t,z)$, is regular at $z=0$. Note that $\bar{\eta}(t,0)$ exists
for all $t\geq 0$ and is a real valued monotone increasing function of $%
t\geq 0$ satisfying $-2\leq \bar{\eta}(t,0)\leq -1$ as the graph of $g(t,p)$
always intercepts the straight line $Q(0,p)=-p/2-1$ at some negative point $%
p^{\ast }(t)$ within that range (see Figure \ref{scpp}) and $p^{\ast }(t)=%
\bar{\eta} (t,0)$ by definition.

It follows that $v(t,\eta )$ is holomorphic in $\Re e\left( \eta \right) <0$
with $v\left( t,\bar{\eta}(t,0)\right) =0$ and 
\begin{equation}
v_{\eta }\left( t,\bar{\eta}(t,0)\right) =\frac{1}{2\bar{\eta}^{2}}+\frac{1}{%
\bar{\eta}^{4}}\ln \left( 1-(e^{2t}-1)\bar{\eta}\right) +\frac{1}{\bar{\eta}%
^{2}}\left( 1+\frac{1}{\bar{\eta}}\right) \frac{e^{2t}-1}{1-(e^{2t}-1)\bar{%
\eta}}>0  \label{veta}
\end{equation}%
and these are the assumptions of our second ingredient (see Theorem $9.4.1$
of \cite{Hi} for a proof)

\begin{theorem}
\label{ift}Let $R>r>0$ and $\bar{\eta}\in \mathbb{C}$ be such that $v(t,\eta
)$ is holomorphic in $D_{R}(\bar{\eta})=\left\{ \eta \in \mathbb{C}%
:\left\vert \eta -\bar{\eta}\right\vert <R\right\} $, $v(t,\bar{\eta})=0$, $%
v_{\eta }\left( t,\bar{\eta}\right) >0$ and $v(t,\eta )\neq 0$ for $%
0<\left\vert \eta -\bar{\eta}\right\vert <r$. Then the contour integral%
\begin{equation*}
\bar{\eta}(t,z):=\frac{1}{2\pi i}\int_{\mathcal{C}}\eta \frac{v_{\eta
}(t,\eta )}{v(t,\eta )-z}~d\eta
\end{equation*}%
where $\mathcal{C}=\left\{ \eta \in \mathbb{C}:\left\vert \eta -\bar{\eta}%
\right\vert =\rho \right\} $ for some $\rho <r$, defines a holomorphic
function in $\left\{ z:\left\vert z\right\vert <m\right\} $ where%
\begin{equation*}
m=\min_{\theta }\left\vert v(t,\bar{\eta}+\rho e^{i\theta })\right\vert ~.
\end{equation*}%
Moreover, $\eta =\bar{\eta}(t,z)$ is the unique solution of $z=v(t,\eta )$
regular at $z=0$ in this domain.
\end{theorem}

For fixed $t$, let $R=R(t)$ be such that $D_{R}(\bar{\eta}(t,0))\subset
\left\{ \Re e\left( \eta \right) <0\right\} $ and note that we can always
take $R$ large enough to include $\eta =-1$. Let $r<R$ be so that $v(t,\eta
)\neq 0$ for $0<\left\vert \eta -\bar{\eta}(t,0)\right\vert <r$. This is
always possible by continuity in view of (\ref{veta}). Finally we pick $\rho
<r$ which gives the largest $m$. As $t$ gets large, $\bar{\eta}(t,0)$
approaches $-1$ and $\rho $ may be chosen so that $m(t)=\min_{\theta
}\left\vert v(t,\bar{\eta}(t,0)+\rho e^{i\theta })\right\vert $ grows like $%
t $, namely, for $\rho $ close to $1/2$. In the limit $t\rightarrow \infty $%
, $\bar{\eta}(t,z)$ becomes holomorphic in the entire complex plane.

\begin{figure}[th]
\begin{center}
\epsfig{file=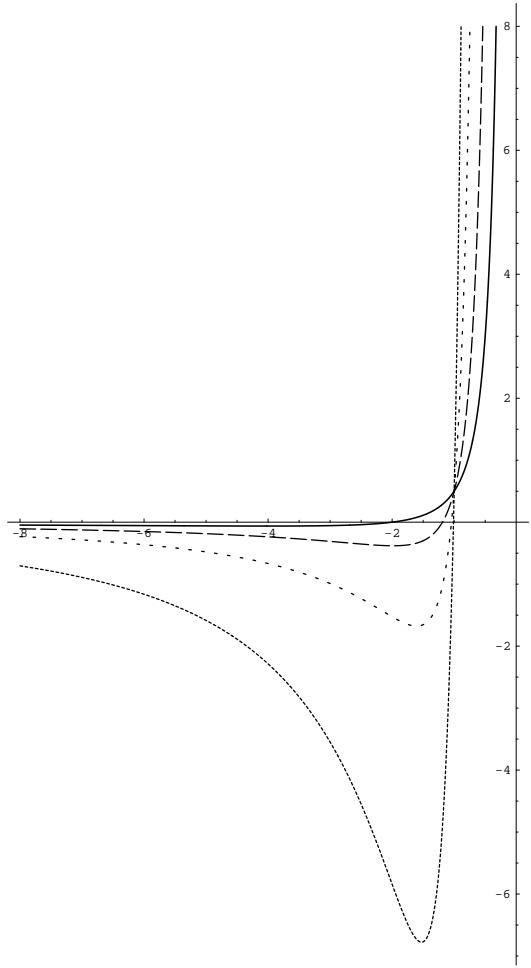, scale=0.5
}
\end{center}
\caption{Profile of (\protect\ref{vtp}) for $t=0$ (solid line), $10$ (long
dashes), $10^{5}$ (short dashes) and $10^{20}$ (dots) and $p$ in a
neighborhood of $p=-1$}
\label{profile}
\end{figure}

To describe the asymptotic behavior of $\bar{\eta}\left( t,z\right) $ as $%
t\rightarrow \infty $, equation (\ref{vx}) can be written as%
\begin{eqnarray*}
v(t,\bar{\eta}) &=&\frac{1}{2\bar{\eta}^{2}}-\frac{\bar{\eta}+1}{\bar{\eta}%
^{3}}\left\{ \frac{-\bar{\eta}}{2}+\ln \left( -\bar{\eta}\right) +2t+\ln
\left( 1-\frac{\bar{\eta}+1}{\bar{\eta}}e^{-2t}\right) \right\} \\
&=&\frac{1}{2}+2t(\bar{\eta}+1)+O\left( t\left( \bar{\eta}+1\right) ^{2},(%
\bar{\eta}+1)\right) =z
\end{eqnarray*}%
which gives%
\begin{equation*}
\bar{\eta}(t,z)=-1-\frac{1}{2t}\left( \frac{1}{2}-z\right) +R\left(
t,z\right)
\end{equation*}%
where, by Theorem \ref{ift}, $R$ is a regular function of $z$ for $%
\left\vert z\right\vert <m(t)$ which goes to $0$ faster than $1/t$,
concluding the proof of Theorem \ref{gauss}. Note that 
\begin{equation*}
\lim_{t\rightarrow \infty }\lim_{N\rightarrow \infty }\frac{1}{N}U(t,\sqrt{N}%
z)=\lim_{t\rightarrow \infty }u(t,x)=\int_{0}^{x}\lim_{t\rightarrow \infty }%
\bar{p}(t,x^{\prime })~dx^{\prime }=-x=\left\vert z\right\vert ^{2}
\end{equation*}%
and $U_{0}(z)=\left\vert z\right\vert ^{2}$ is an equilibrium solution of (%
\ref{V}), for any number of components $N$.

\hfill $\Box $

\begin{remark}
\label{multivalue}Figure \ref{profile} shows the solution $v(t,p)$ of (\ref%
{v}), for various $t$. For $t=0$, $v(0,p)=v_{0}(p)$ is a monotone increasing
(decreasing) function of $p\in \left( -4,0\right) $ ($p\in \left( -\infty
,-4\right) $) and its inverse $v_{0}^{-1}(x)=u_{0}^{\prime }(x)$ is defined
for $x\in \left( -1/16,\infty \right) $. For $t>0$, there is a unique
negative value $-l(t)$ (with $l(0)=4$), given by $v_{p}(t,-l(t))=0$, such
that $v(t,p)$ is monotone decreasing if $-\infty <p<-l(t)$ and monotone
increasing if $-l(t)\leq p<0$. The inverse function $v^{-1}(t,x)=u_{x}(t,x)$
has two branches but only the one with $v^{-1}(0,x)=u_{0}^{\prime }(x)$
converges to $-1$ in any compact interval inside $\left( -d(t),\infty
\right) $ with $-d(t)=v(t,-l(t))<-1/16$ for all $t>0$ and $d(t)\rightarrow
\infty $ as $t\rightarrow \infty $.
\end{remark}

\section{Geometry of the Scaling Flow \label{GFT}}

\setcounter{equation}{0} \setcounter{theorem}{0}

\noindent \textbf{Critical Trajectory.} The scaling flow $u(t,x)$, defined
by equations (\ref{U}) and (\ref{uV}), is the cummulant generating function
of the block spin variable at scale $t$. The flow is determined by its
partial derivative $u_{x}(t,x)$ (see (\ref{up})) and Theorem \ref{gauss}
exhibits a single trajectory, in the (viscosity) limit $N\rightarrow \infty $%
, 
\begin{equation*}
\mathcal{O}(u_{0}^{\prime }\rightarrow -1)=\left\{
u_{x}(t,x),~t>0:~u_{x}(0,x)=u_{0}^{\prime }(x),~u_{x}(\infty ,x)\equiv
-1\right\} ,
\end{equation*}%
that starts at $t=0$ from the initial function (\ref{pux}) and converges, as 
$t$ goes to $\infty $, to the stationary solution $-1$, implicitly defined
by (\ref{vx}) and (\ref{pbar}). In this subsection we identify the class of
functions where the flow is defined and give a geometric function theory
description of this trajectory that establishes a one-to-one and onto
relation between the orbit $\mathcal{O}(u_{0}^{\prime }\rightarrow -1)$ and
the time dependent convex domains $\Omega (t)=u_{x}(t,\mathbb{H})$, $t\geq 0$%
, formed by images under $u_{x}$ of the upper half--plane $\mathbb{H}%
=\left\{ z=x+iy\in \mathbb{C}:y>0\right\} $. Analytical and numerical
techniques are combined in order the conformal equivalence between $\Omega
(t)$ and $\mathbb{H}$ to be explicitly verified for all $t$.

\medskip

\noindent \textbf{Analytic Continuation of initial value.} Let us begin by
extending Watanabe's proof of Proposition \ref{visclim} to the upper
half--plane $\mathbb{H}$.

\medskip

\noindent \textit{Proof of Proposition \ref{visclim}.} Let $\phi _{\nu }(\xi
)=\xi J_{\nu }(\xi )/J_{\nu -1}(\xi )$ be defined for $\nu \geq 1$ and $\xi
\in \mathbb{C}$. The Bessel recursion relation%
\begin{equation*}
J_{\nu -1}(\xi )+J_{\nu +1}(\xi )=\frac{2\nu }{\xi }J_{\nu }(\xi )
\end{equation*}%
generates a continued fraction of Gauss (see Chapter $X\!V\!I\!I\!I$ of \cite%
{Wa}): 
\begin{equation}
\phi _{\nu }(\xi )=\dfrac{2}{\nu }\frac{\left( \xi /2\right) ^{2}}{1-\dfrac{1%
}{2\nu }\phi _{\nu +1}(\xi )}=\dfrac{2}{\nu }\frac{\left( \xi /2\right) ^{2}%
}{1-\dfrac{1}{\nu (\nu +1)}\dfrac{\left( \xi /2\right) ^{2}}{1-\dfrac{1}{%
2\nu +2}\phi _{\nu +2}(\xi )}}  \label{phinu}
\end{equation}%
uniformly convergent over the domain 
\begin{equation}
\dfrac{1}{\nu (\nu +1)}\left\vert \xi \right\vert ^{2}\leq 1~,  \label{D}
\end{equation}%
by Worpitzky's Theorem (see \cite{Wa}, p. $42$).

Let $\vartheta _{N}(x):=U(0,\sqrt{N}z)/N$ with $x=-\left\vert z\right\vert
^{2}$. Equation (\ref{V0}) together with (\ref{ic}) and the Bessel recursion
relation $\nu J_{\nu }(\xi )-\xi J_{\nu }^{\prime }(\xi )=\xi J_{\nu +1}(\xi
)$, gives%
\begin{eqnarray}
x\vartheta _{N}^{\prime }(x) &=&\frac{1}{2N}\left\{ \frac{(N/2-1)J_{N/2-1}(i%
\sqrt{\beta x}N)-i\sqrt{\beta x}N~J_{N/2-1}^{\prime }(i\sqrt{\beta x}N)}{%
J_{N/2-1}(i\sqrt{\beta x}N)}\right\}  \notag \\
&=&\frac{1}{2N}\phi _{N/2}(i\sqrt{\beta x}N)~.  \label{xtheta}
\end{eqnarray}%
We take $\xi =i\sqrt{\beta x}N$ and $\nu =N/2$ in (\ref{phinu}) and write%
\begin{equation*}
\frac{1}{2N}\phi _{N/2}(i\sqrt{\beta x}N)=\frac{-1}{2}\frac{a_{0}}{1-\dfrac{%
a_{1}}{1-\dfrac{a_{2}}{1-\ddots }}}~.
\end{equation*}%
As $N$ goes to infinity,%
\begin{equation*}
a_{k}=\frac{-\beta x}{\left( 1+\dfrac{2k}{N}\right) \left( 1+\dfrac{2k+2}{N}%
\right) }
\end{equation*}%
converges to $-\beta x$ uniformly over the domain (\ref{D}) for any integer $%
k\geq 0$ and, consequently, $x\vartheta _{N}^{\prime }(x)$ converges over
the same domain to a periodic continued fraction. We thus have%
\begin{equation}
xu_{0}^{\prime }(x)=\lim_{N\rightarrow \infty }x\vartheta _{N}^{\prime }(x)=%
\frac{-1}{2}\frac{\beta x}{1+\dfrac{\beta x}{1+\dfrac{\beta x}{1+\ddots }}}=%
\frac{-\beta x}{1+\sqrt{1+4\beta x}}~~  \label{pick}
\end{equation}%
where the third equality is $-1/2$ times the solution $\phi $ of 
\begin{equation*}
\phi =\frac{\beta x}{1+\phi }
\end{equation*}%
that is positive for positive $x$. This yields (\ref{u0}) in view of the
normalization $\vartheta _{N}(0)=0$. Note that the limit holds for any $x$
in the domain%
\begin{equation}
\left\vert 4\beta x\right\vert \leq 1  \label{D1}
\end{equation}%
of complex plane and this is sharp for the limit function $u_{0}$ since $%
(-\infty ,-1/4\beta ]$ is a branching cut of $u_{0}^{\prime }$.

\medskip

\noindent \textbf{Lee--Yang zeroes.} Both functions $\vartheta _{N}^{\prime
}(x)$ and $u_{0}^{\prime }(x)$ can be analytic continued to the upper
half--plane and extended, by reflection, to the slit domain $\mathbb{C}%
\backslash (-\infty ,-1/4\beta ]$. As $\phi _{\nu }(\xi )$ is a meromorphic
(even) function of $\xi $, it can be written as%
\begin{equation}
\phi _{\nu }(\xi )=\xi ^{2}\sum_{n\geq 1}\frac{1}{\alpha _{n,\nu -1}^{2}-\xi
^{2}}  \label{phinuzeta}
\end{equation}%
where $\alpha _{n,\nu }$, $n\geq 1$, are zeroes of the Bessel function $%
J_{\nu }$. So, the limit $N\rightarrow \infty $ of (\ref{xtheta}) together
with the asymptotic behavior of the Bessel's zeroes, 
\begin{equation*}
\alpha _{n,N/2-1}\sim \left( N-1\right) \dfrac{\pi }{4}+\left( 2n-1\right) 
\dfrac{\pi }{2}
\end{equation*}%
for $n$ large, gives 
\begin{equation}
u_{0}^{\prime }(x)=\lim_{N\rightarrow \infty }\vartheta _{N}^{\prime
}(x)=\lim_{N\rightarrow \infty }\frac{1}{2N}\sum_{n\geq 1}\frac{-\beta }{%
\dfrac{\alpha _{n,N/2-1}^{2}}{N^{2}}+\beta x}=\frac{1}{2\pi }%
\int_{1/4}^{\infty }\frac{\beta }{-g(s)-\beta x}ds  \label{lyzeros}
\end{equation}%
for some positive function $g$ satisfying $g(s)\sim s^{2}$ for large $s$.
Note that $\left\{ \alpha _{n,N/2-1},~n\geq 1\right\} $ are the Lee--Yang
zeroes of the \textquotedblleft a priori\textquotedblright\ initial measure (%
\ref{ic}) and, by (\ref{V0}) and (\ref{uU}), they become dense over an
interval of real line.

\medskip

\noindent \textbf{Pick class of functions.} Let $P$ denote the class of
functions 
\begin{equation*}
f(\zeta )=u(\zeta )+iv(\zeta )\ ,\qquad \zeta =x+iy,
\end{equation*}%
analytic in the upper half--plane $\mathbb{H}$ with positive imaginary part: 
$v(\zeta )\geq 0$ if $y>0$ (see e.g. \cite{Do}, Chap. $I\!I$). The class of
functions $P$ forms a convex cone and is closed under composition:

\begin{enumerate}
\item $af_{1}+bf_{2}\in P$

\item $f_{1}\circ f_{2}\in P$
\end{enumerate}

\noindent hold for any $a,b\geq 0$ and $f_{1},f_{2}\in P$.

A linear function $a+b\zeta $, $a\in \mathbb{R}$ and $b>0$, and the function 
$-1/\zeta $ are clearly in $P$ since both are one-to-one and onto maps of $%
\mathbb{H}$ into itself. It thus follows by (\ref{phinuzeta}), together with
the properties $1.$ and $2.$, that $\phi _{\nu }(\zeta )$ is in $P$ and, in
the topology of uniform convergence on compact subsets of $\mathbb{H}$, the
sequence $\left( \vartheta _{N}^{\prime }\right) _{N\geq 1}$ converges to $%
\vartheta _{\infty }^{\prime }$ in $P$ (\cite{Do}, Sec. $4$ in Chap. $I\!I$%
). Note the following equality $\vartheta _{\infty }^{\prime
}(x)=u_{0}^{\prime }(x)$ in the domain (\ref{D1}) and $u_{0}^{\prime }$ is
the composition of four Pick functions: $1+\beta \zeta $, $\sqrt{\zeta }$, $%
\left( 1+\zeta \right) /\beta $ and $-1/\zeta $. This implies that $%
u_{0}^{\prime }\in P$ and equality between first and last expression in (\ref%
{pick}) holds with $x$ replaced by $\zeta \in \mathbb{H}$, concluding the
proof of Proposition \ref{visclim}.

\hfill $\Box $

\medskip

\noindent \textbf{Integral representation.} A function $f(\zeta )=u\left(
\zeta \right) +iv\left( \zeta \right) $ is in the Pick class if and only if
has a unique canonical integral representation \cite{Do}%
\begin{equation}
f(\zeta )=a~\zeta +b+\int_{-\infty }^{\infty }\left( \frac{1}{\lambda -\zeta 
}-\frac{\lambda }{\lambda ^{2}+1}\right) d\mu (\lambda )
\label{representation}
\end{equation}%
where $a=\lim\limits_{y\rightarrow \infty }f\left( iy\right) /iy\geq 0$, $%
b=u(i)$ is real and $\mu $ is a positive Borel measure on $\mathbb{R}$ such
that $\displaystyle\int \left( \lambda ^{2}+1\right) ^{-1}~d\mu (\lambda
)<\infty $. In addition,%
\begin{equation}
\mu \left( (a,b)\right) +\frac{\mu \left( \{a\}\right) +\mu \left(
\{b\}\right) }{2}=\lim_{y\downarrow 0}\frac{1}{\pi }\int_{a}^{b}v(x+iy)~dx
\label{mu}
\end{equation}%
holds for any finite interval $\left( a,b\right) $ and determines $\mu $
uniquely from $f$.

The initial condition $u_{0}^{\prime }(x)$ of the flow $u_{x}(t,x)$ goes to $%
0$ as $x$ goes to infinity (in any direction of the complex plane).
Consequently, $a$ of its canonical representation vanishes. In addition, $b$
can be identified with the second integral. So, if%
\begin{equation}
f_{0}(\zeta )=\int_{-\infty }^{\infty }\frac{1}{\lambda -\zeta }d\mu \left(
\lambda \right)  \label{f0}
\end{equation}%
is defined with $d\mu (\lambda )=\rho (\lambda )d\lambda $ an absolutely
continuous measure w.r.t. the Lebesgue measure $d\lambda $:%
\begin{equation}
\rho (\lambda )=\frac{1}{4\pi }\frac{\sqrt{4\left( -\lambda \right) -1}}{%
\left( -\lambda \right) }  \label{rho0}
\end{equation}%
whose support is $-\infty <\lambda <-1/4$, then%
\begin{equation*}
u_{0}^{\prime }(x)=\beta f_{0}(\beta x)
\end{equation*}%
by (\ref{mu}). Note that $\displaystyle\int_{-\infty }^{\infty }\lambda
^{-1}d\mu \left( \lambda \right) =-1/2$ agrees with $f_{0}(0)$, by an
explicit integration. Equation (\ref{f0}) together with (\ref{lyzeros})
leads to the following relation between (\ref{rho0}) and the empirical
density $\sqrt{g(s)}$ of Lee--Yang zeroes: $2\pi \rho (\lambda )=\left[
\left( -g^{\prime }\circ g^{-1}\right) (\lambda )\right] ^{-1}$.

\medskip

\noindent \textbf{Geometric function theory.} By the Riemann mapping theorem
(see e.g. \cite{GK}) if an open set $\Omega $ is topologically equivalent to 
$\mathbb{H}$ (i. e. $\Omega $ and $\mathbb{H}$ are homeomorphic) then $%
\Omega $ is also conformally equivalent to $\mathbb{H}$ and there exist a
biholomorphic (holomorphic one-to-one and onto) mapping $f$ from $\Omega $
to $\mathbb{H}$. In some cases $f$ can be made uniquely defined by $\Omega $%
. The conformal equivalence of open sets provides qualitative informations
on the trajectory $\mathcal{O}(u_{0}^{\prime }\rightarrow -1)$. As a
function of the Pick class, $u_{0}^{\prime }$ maps $\mathbb{H}$ into itself
but we can be more specific about the image of $\mathbb{H}$ by $%
u_{0}^{\prime }$. From here on we fix $\beta $ at the critical value $\beta
_{c}(4)=4$. We denote the upper semi--disc of radius $r$ centered at $x_{0}$
by%
\begin{equation*}
\mathbb{S}_{r}(x_{0})=\left\{ \zeta =x+iy\in \mathbb{H}:\left(
x-x_{0}\right) ^{2}+y^{2}<r^{2}~\right\} ~
\end{equation*}%
and let $\mathcal{S}_{t}$ be the class in $P$ indexed by $t\in \mathbb{R}%
_{+} $ satisfying

\begin{enumerate}
\item[(i)] $\varphi $ is an univalent function (one-to-one)

\item[(ii)] $\varphi (\zeta _{t})=\zeta _{t}$ for some complex number $\zeta
_{t}$

\item[(iii)] $\varphi (1/2)=-1$

\item[(iv)] $\varphi (\bar{\zeta})=\bar{\varphi}(\zeta )$
\end{enumerate}

\begin{proposition}
\label{domain0}$u_{0}^{\prime }$ maps the upper half--plane $\mathbb{H}$
conformally into the interior of the upper semi--disc of radius $2$ centered
at $-2$: 
\begin{equation*}
u_{0}^{\prime }(\mathbb{H})=\Omega _{0}=\mathbb{S}_{2}(-2)~
\end{equation*}%
and no other function in $\mathcal{S}_{0}$ maps $\mathbb{H}$ into $\Omega
_{0}$. Hence, there is a one--to--one and onto relation between $\Omega _{0}$
and the initial function $u_{0}^{\prime }$ of critical trajectory $\mathcal{O%
}(u_{0}^{\prime }\rightarrow -1)$ in the class $\mathcal{S}_{0}$ of
functions with fixed point $\zeta _{0}$ given by the complex root of $%
2x^{3}-x-2$.
\end{proposition}

\noindent \textit{Proof}. By equations (\ref{vx}) and (\ref{p}), the inverse
of $u_{0}^{\prime }$, given by 
\begin{equation}
v_{0}(p)=\frac{p+2}{2p^{2}}~,  \label{pp}
\end{equation}%
is the initial condition (\ref{v0}) of the linear evolution equation (\ref{v}%
). Hence%
\begin{equation}
\Omega _{0}=\left\{ \eta =p+iq\in \mathbb{H}:\Im \left( v_{0}(p+ip)\right)
>0\right\} ~  \label{omega0}
\end{equation}%
and this is equivalent, by (\ref{pp}), to the following inequalities 
\begin{equation*}
q\left( p^{2}-q^{2}\right) -2pq\left( p+2\right) >0,\qquad q>0
\end{equation*}%
which can be written as the upper semi--disc $\mathbb{S}_{2}^{+}(-2)$: $%
\left( p+2\right) ^{2}+q^{2}<4$, $q>0$.

Since%
\begin{equation*}
v_{0}^{\prime }(p)=-\frac{p+4}{2p^{3}}
\end{equation*}%
does not vanish neither diverges for any $p$ in $\Omega _{0}$ but at edge
points $p=-4$ and $p=0$ in the closure $\overline{\Omega _{0}}$ of $\Omega
_{0}$, we conclude by (\ref{pp}) that $u_{0}^{\prime }\left( \mathbb{H}%
\right) =\Omega _{0}$ is one--to--one and onto map. Note that $u_{0}^{\prime
\prime }(x)$ vanishes at $x=v_{0}(p_{\infty })$ with $p_{\infty }$ such that 
$v_{0}^{\prime }(p_{\infty })=\infty $, in view of $u_{0}^{\prime \prime
}\circ v_{0}(p)=1/v_{0}^{\prime }(p)$, i.e., at infinity in every direction
of the complex plane.

Now, suppose there exist another function $\varphi (x)$ in $\mathcal{S}_{0}$
such that $\varphi (\mathbb{H})=\Omega _{0}$. Then, $\varphi ^{-1}\circ
u_{x} $ is a map from $\mathbb{H}$ onto itself, belongs to the class $P$ and
leave the points $1/2$, $\zeta _{0}$ and $\overline{\zeta _{0}}$ fixed.%
\footnote{%
The class of functions in $P$ considered can be analytically continued
across the real line by reflection (see condition $(iv)$ of $\mathcal{S}_{t}$%
). If $\zeta _{0}=x_{0}+iy_{0}\in \mathbb{H}$ is a fixed point of $f\in P$
then $\overline{\zeta _{0}}=x_{0}-iy_{0}$ is a fixed point of its extension.}
As a consequence of (\ref{representation}), the Pick functions that map $%
\mathbb{H}$ onto $\mathbb{H}$ are linear fraction transformations. Since the
identity mapping is the only linear fraction transformation leaving three
points fixed, we infer that $\varphi (x)$ and $u_{x}(x)$ are the same
function. The complex root $\zeta _{0}\simeq -0.582687+0.720119i$ of the
fixed point equation%
\begin{equation*}
v_{0}(\zeta )=\frac{\zeta +2}{2\zeta ^{2}}=\zeta ~
\end{equation*}%
is in $\Omega _{0}$, concluding the proof of Proposition \ref{domain0}.

\hfill $\Box $

\medskip

We now apply (\ref{omega0}) to determine $\Omega _{t}=u_{x}\left( t,\mathbb{H%
}\right) $ for $t>0$. As $u_{x}(t,\zeta )$ solves $v(t,\eta )=\zeta $ for $%
\eta $, with $v$ explicitly given by (\ref{vtp}), the domain $\Omega _{t}$
can be easily plotted using ContourPlot or ImplicitPlot packages in \textit{%
Mathematica}. Approximate expressions can be given for $t$ around $0$ and $%
\infty $.

\medskip

\begin{figure}[th]
\begin{center}
\epsfig{file=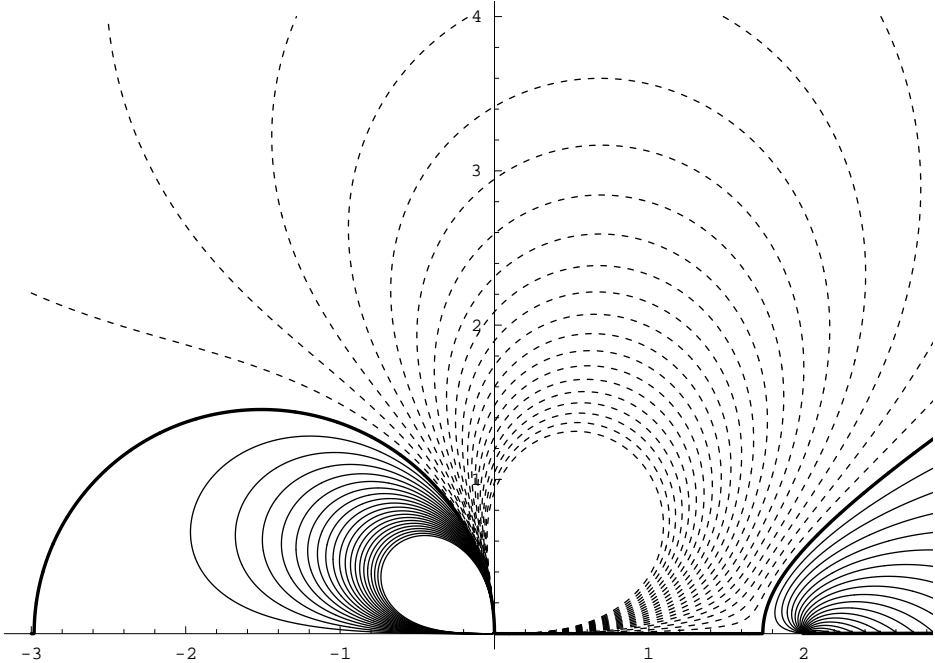, scale=0.7
}
\end{center}
\caption{$\Im \left( v(t,p+iq)\right)= c $ for $t=0.2$ with $c$ taking
negative (dashed lines), positive (solid curves) and neutral (thick solid
line) values}
\label{contour}
\end{figure}

\noindent \textbf{Domain boundary.} Each set $\Omega _{t}$ of the family for 
$t>0$ is bounded by a simple convex closed curve which is piecewise analytic
and defined by equation 
\begin{equation}
\Im \left( v(t,\eta )\right) =0,\ \eta =p+iq\in \mathbb{\bar{H}}
\label{zero}
\end{equation}%
where $v(t,\eta )$ is analytically continued to the closure $\mathbb{\bar{H}}
$ of the half--plane $\mathbb{H}$. One has to be careful, however, in order
to get the actual domain since $\Im \left( v(t,p+iq)\right) >0$ may have
more than one component. Figure \ref{contour} shows level curves of $\Im
\left( v(t,p+iq)\right) $.

For $t$ small, $\Omega _{t}$ is a slight deformation of $\mathbb{S}_{2}(-2)$%
, by continuity: 
\begin{equation*}
\left( p+\frac{2(1+2t)}{1+4t}\right) ^{2}+q^{2}<\frac{4(1+2t)^{2}}{\left(
1+4t\right) ^{2}}~,\;q>0~.
\end{equation*}%
Whereas, for $t$ very large, $\Omega _{t}$ approaches\ a folium (half--leaf)
of Decartes:%
\begin{equation*}
\Im \left( 2t\frac{1+p+iq}{\left( p+iq\right) ^{3}}\right) \geq 0,\
q>0\Longleftrightarrow 2p\left( p^{2}+q^{2}\right) +3p^{2}-q^{2}\leq 0\,,\
q>0~.
\end{equation*}

\begin{figure}[th]
\begin{center}
\epsfig{file=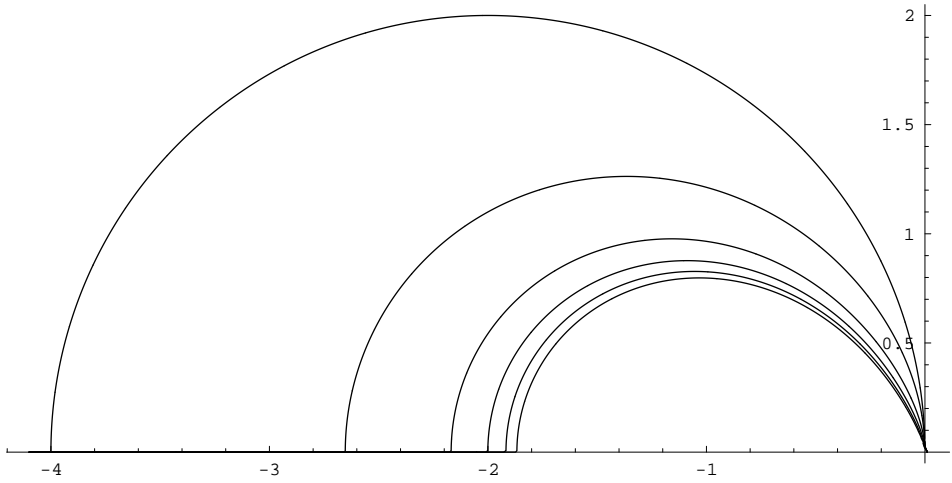, scale=0.7 
}
\end{center}
\caption{Domain boundaries $\Omega _{t}$ for $t=n/4$, $n=0,\ldots ,9 $}
\label{semicircles}
\end{figure}

We observe that the boundary of $\Omega _{t}$ is the union of two curves: a
line segment $I_{\alpha }:=[-\alpha ,0]$ extending from a point $-\alpha
=-\alpha (t)<0$ up to the origin over the real line and a convex\ curve $%
q=h(t,p)$ defined for $p\in I_{\alpha }$ with $h(t,-\alpha )=h(t,0)=0$. From
the above, $\alpha (t)$ is a monotone decreasing function of $t$ with $%
\alpha (0)=4$ and $\lim_{t\rightarrow \infty }\alpha (t)=3/2$ whereas $%
h(t,p) $ is a semi--circular curve at $t=0$: $h(0,p)=\sqrt{4-\left(
p+2\right) ^{2}} $ and approaches a limit (half--leaf) curve 
\begin{equation*}
h^{\ast }(p)=\lim_{t\rightarrow \infty }h(t,p)=\sqrt{\frac{3p^{2}+2p^{3}}{%
1-2p}}.
\end{equation*}%
Figure \ref{semicircles} shows domain boundaries for various $t$. Note that $%
\Omega _{t}\subset \Omega _{t^{\prime }}$ if $t^{\prime }<t$ with strict
inclusion along the convex arc.

The function $h(0,p)$ and the turning point $-\alpha (0)=-4$ are related to
the density $\rho (\lambda )$ of the canonical representation (\ref{f0}) of $%
u_{0}^{\prime }(x)$ and its support $\Sigma _{0}$, 
\begin{equation*}
\rho \left( \frac{1}{p}\right) =\frac{1}{4\pi }\sqrt{-4p-p^{2}}=\frac{1}{%
4\pi }h(0,p)~,\ -4\leq p\leq 0
\end{equation*}%
by substituting $\lambda =1/p$ in (\ref{rho0}). Note that $\Sigma
_{0}=\Sigma (0)=\left( -\infty ,-d(0)\right) $ in this case is such that $%
d(0)=\left( 4\alpha (0)\right) ^{-1}=1/16$.

To determine the support $\Sigma (t)=\left( -\infty ,-d(t)\right) $ of the
measure $\mu (t,d\lambda )=\rho (t,\lambda )d\lambda $ of the canonical
representation of $u_{x}(t,x)$ we look at the negative value $-l(t)$ at
which $v_{p}(t,-l(t))=0$. In the neighborhood of this point $v$ is not
univalent. Observe that $l(t)$ and the turning point $\alpha (t)$ coincide.
Writing 
\begin{equation}
v(t,\eta )=y(t,p,q)+iw(t,p,q),\qquad \eta =p+iq,  \label{vyw}
\end{equation}%
by definition of $\alpha $ and Cauchy--Riemann equations, we have%
\begin{equation*}
0=w_{q}\left( t,-\alpha (t),0\right) =y_{p}\left( t,-\alpha (t),0\right)
=v_{p}(t,-\alpha (t))
\end{equation*}%
which implies $\alpha (t)=l(t)$ by uniqueness. From Remark \ref{multivalue},
we have $-d(t)=v(t,-l(t))<-1/16$ for all $t>0$ and\ 
\begin{eqnarray*}
-d(t) &\sim &v(t,-3/2) \\
&=&\frac{1}{9}-\frac{4}{27}\ln \left( 1+\frac{3}{2}(e^{2t}-1)\right) =\frac{%
-8}{27}t+O(1)~,
\end{eqnarray*}%
for $t$ large enough, implies that the support $\Sigma (t)$ of $\mu
(t,\lambda )$ converges to an empty set: $\Sigma (t)=\left( -\infty
,-d(t)\right) \rightarrow \emptyset $ as $t\rightarrow \infty $.

\medskip

\begin{figure}[th]
\begin{center}
\epsfig{file=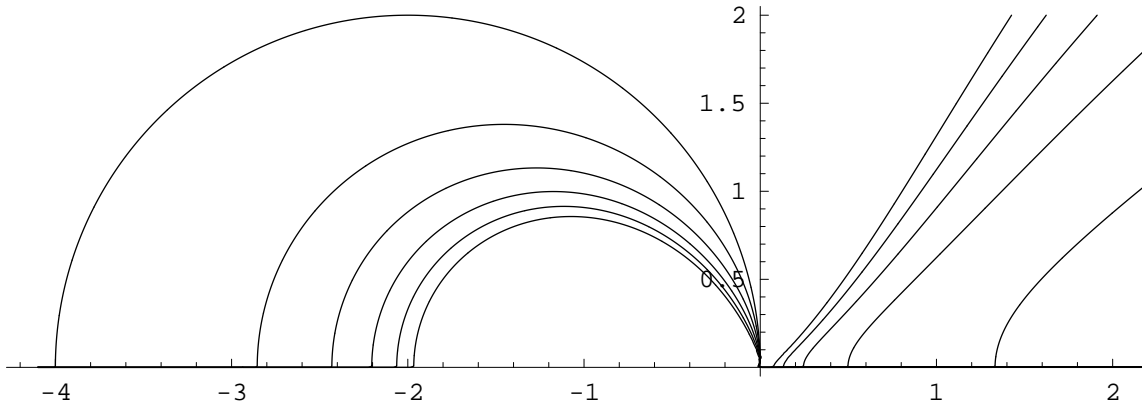, scale=1
}
\end{center}
\caption{Domain boundaries $\Omega _{t}$ and $\Lambda _{t}$ for $t=n/4$, $%
n=0,\ldots ,9 $}
\label{semicircles2}
\end{figure}

\noindent \textbf{Riemann surfaces.} Contour plots of $v(t,\eta )$, $\eta
\in \mathbb{C}$, for various $t$, show that $u_{x}(t,\zeta )$ is a
multivalued function of $\zeta \in $ $\mathbb{C}$. Already at $t=0$, $%
u_{0}^{\prime }(\zeta )$ has two Riemann surfaces connected by a branch cut
along the segment $(-\infty ,-1/16]$ across which the imaginary part of $%
u_{0}^{\prime }(\zeta )$ change sign ($u_{0}^{\prime }((-\infty ,-1/16])$ is
the semi--circular boundary of $\mathbb{S}_{2}(-2)$). The determination of $%
\sqrt{\cdot }$ is chosen such that $-1/\left( 1+\sqrt{1+16\zeta }\right) $
is in $P$. For $t>0$, $u_{x}(t,\zeta )$ has an even more elaborate Riemann
surface with three sheets. The first is connected with the second sheet by a
branch cut $(-\infty ,-d(t)]$ while the latter is also connected to a third
sheet by a branch cut $(0,d_{1}(t)]$ with $d_{1}(0)=0$ and $%
d_{1}(t)\rightarrow \infty $ as $t\rightarrow \infty $, which does not
concern us as it doesn't relate to the limit function $-1$. The curves $%
u_{x}(t,(-\infty ,-d(t)])$ and $u_{x}(t,(0,d_{1}(t)])$, which define
together with the real line boundaries of two domains, intercept the real
line perpendicularly at negative and positive values, respectively. Figure %
\ref{semicircles2} shows these curves for various $t$. The region bounded by 
$u_{x}(t,(0,d_{1}(t)])$ inside the half--plane $\mathbb{H}$ is denoted by $%
\Lambda _{t}$. Note that, opposed to $\Omega _{t}$, $\Lambda _{t}$ are open
domains satisfying inclusions $\Lambda _{t}\subset \Lambda _{t^{\prime }}$
if $t<t^{\prime }$.

\medskip

\noindent \textbf{Flow in the Pick class.} It is very difficult to show
directly from the flow equation that $u_{x}(t,\zeta )$ remains in the Pick
class of functions for all $t>0$ by general principles. However, for initial
condition in $P$ that belongs to the class $\mathcal{S}_{0}$ there is a
simple property of the flow equation that explains why the Pick class $P$ is
preserved. Writing $v_{0}(\eta )=y_{0}(p,q)+iw_{0}(p,q)$ as a function of $%
\eta =p+iq\in \Omega _{0}\cup \Omega _{0}^{\ast }\cup I_{\alpha (0)}$, with $%
\Omega ^{\ast }$ the reflection of $\Omega $ about the real axis, if the
imaginary part $w_{0}(p,q)$ is an odd function of $q$ then the flow equation
(\ref{v}) preserves this property. Writing $v(t,\eta )$ as (\ref{vyw}), we
have 
\begin{equation*}
w(t,p,q)=-w(t,p,-q)
\end{equation*}%
holds for all $t\geq 0$ and $\eta \in \Omega _{t}$. By continuity, it
follows that $\Omega _{t}\subset \mathbb{H}$ and that $v(t,\eta )$ remains a
one--to--one and onto map from $\Omega _{t}$ to $\mathbb{H}$ and these imply
that $u_{x}\left( t,\zeta \right) $ belongs to the class $\mathcal{S}_{t}$
in\ $P$.

\begin{figure}[th]
\begin{center}
\epsfig{file=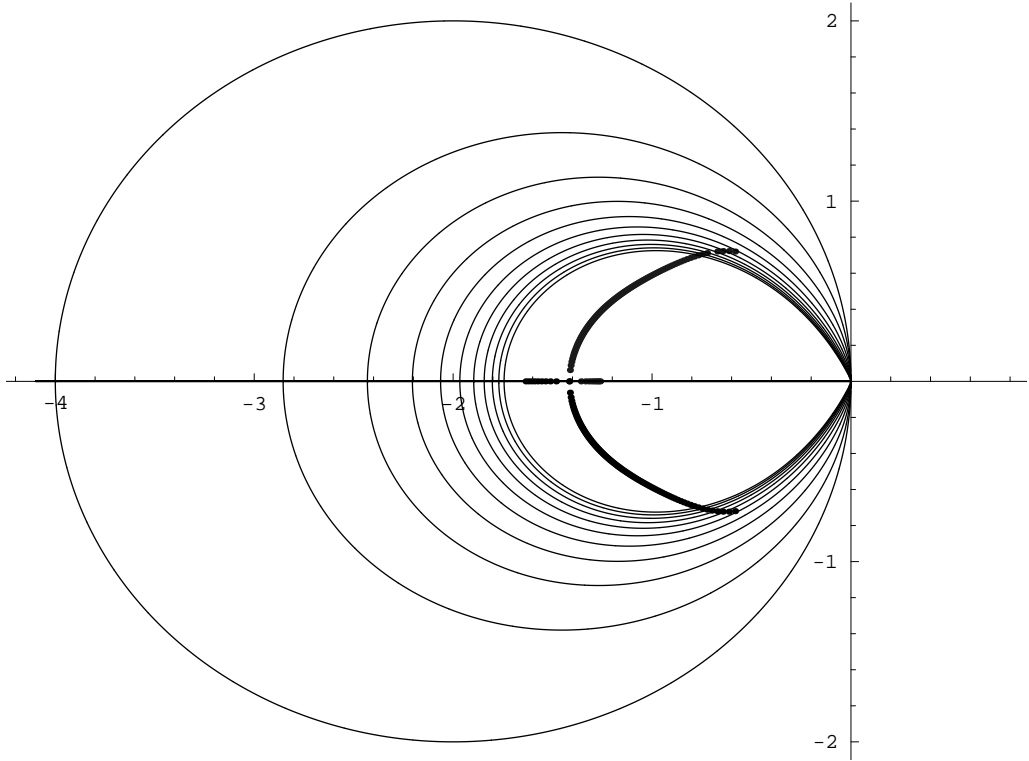, scale=0.8
}
\end{center}
\caption{Scaling evolution of fixed points and domain images of $u_{x}\left(
t,\protect\zeta \right) $}
\label{fixedpoints}
\end{figure}

To establish uniqueness of the relationship between $\mathcal{O}%
(u_{0}^{\prime }\rightarrow -1)$ and the image domains $\left\{ \Omega
_{t},~t\geq 0\right\} $, we proceed as in $t=0$ (see proof of Proposition %
\ref{domain0}). Supposing that $\varphi (t,\zeta )\in \mathcal{S}_{t}$ is a
different function satisfying $\varphi (t,\mathbb{H})=\Omega _{t}$, for each 
$t$ fixed $\varphi ^{-1}\circ u_{x}\left( t,\zeta \right) $ maps $\mathbb{H}$
into itself and leaves the point $1/2$, $\zeta _{t}$ and $\zeta _{t}^{\ast }$
fixed where $\zeta _{t}^{\ast }=\bar{\zeta}_{t}$ for $t<t^{\ast }\simeq
5.155075$ and for $t\geq t^{\ast }$ the last two fixed points become real
numbers (see Figure \ref{fixedpoints}). Extending the functions in $\mathcal{%
S}_{t}$ across the real line by reflection, $\varphi ^{-1}\circ u_{x}\left(
t,\zeta \right) $ is a linear fraction map with three fixed points which
contradicts the hypothesis that $u_{x}\left( t,\zeta \right) $ and $\varphi
(t,\zeta )$ are different. This holds for all $t$ such that $\zeta
_{t}^{\ast }\in \Omega _{t}\cup \Omega _{t}^{\ast }\cup I_{\alpha (t)}$. If
this condition is not satisfied, we apply Schwarzian reflection \cite{Da}
about the curve $h(p)$ in order to extend $u_{x}\left( t,\zeta \right) $ to
the complex plane in such way that $u_{x}\left( t,\mathbb{H}\right) =\Omega
_{t}$ and $u_{x}\left( t,-\mathbb{H}\right) =\mathbb{H}\backslash (\Omega
_{t}\cup \Lambda _{t})$ and this insures that $\zeta _{t}$ and $\zeta
_{t}^{\ast }$, which are now real values, remain fixed points of $%
u_{x}\left( t,\zeta \right) $ when $\zeta _{t}^{\ast }<-\alpha (t)$. The
value $t_{\mathrm{co}}$ that $\zeta _{t_{\mathrm{co}}}^{\ast }=-\alpha (t_{%
\mathrm{co}})$ is called crossover scale from strong to weak (coupling)
regime, term introduced in \cite{HHW}.

The canonical representation of $u_{x}\left( t,\zeta \right) $ is not
suitable for describing the trajectory $\mathcal{O}(u_{0}^{\prime
}\rightarrow -1)$. From the characteristic equations (\ref{pt}) of (\ref{v})
one find that the point $\left( p,V\right) =\left( -1,1/2\right) $ is a
critical point for the two--dimensional dynamical system:%
\begin{equation}
\left( \dot{p},\dot{V}\right) =(-2p\left( 1+p\right)
,-1+(6+4p)V):=(F_{1}(p,V),F_{2}(p,V))  \label{dyn}
\end{equation}%
with $F_{1}(-1,1/2)=F_{2}(-1,1/2)=0$. As $\left( -1,1/2\right) $ is an
invariant point we have $v(t,-1)=1/2$ and, accordingly, $u_{x}\left(
t,1/2\right) =-1$. Instead of fixing $b$ in the canonical representation (%
\ref{representation}) the value of $f$ at $\zeta =i$, we write%
\begin{equation}
u_{x}\left( t,\zeta \right) =-1+\int_{-\infty }^{\infty }\left( \frac{1}{%
\lambda -\zeta }-\frac{1}{\lambda -1/2}\right) d\mu (t,\lambda )~.
\label{ux}
\end{equation}%
Note that, by hypothesis, $\int_{-\infty }^{\infty }\left[ \left( \lambda
-\zeta \right) \left( \lambda -1/2\right) \right] ^{-1}d\mu (t,\lambda
)<\infty $ and as the support $\Sigma (t)=\left( -\infty ,-d(t)\right) $ of $%
\mu \left( t,\lambda \right) $ converges to $\emptyset $ the integral in (%
\ref{ux}) converges to $0$ uniformly in each compact set $O\in \mathbb{H}$.

The following summarizes our findings.

\begin{proposition}
\label{conformal}$u_{x}\left( t,\zeta \right) $, $t>0$, map the upper
half--plane $\mathbb{H}$ conformally into a decreasing family of open convex
sets $\Omega _{t}$ satisfying 
\begin{equation*}
\Omega _{t}=u_{x}\left( t,\mathbb{H}\right) \subset u_{0}^{\prime }(\mathbb{H%
})=\Omega _{0}~
\end{equation*}%
and no other function in $\mathcal{S}_{t}$ maps $\mathbb{H}$ into $\Omega
_{t}$. There is a one--to--one and onto relation between this family and the
trajectory $\mathcal{O}(u_{0}^{\prime }\rightarrow -1)$ at the critical
inverse temperature $\beta =\beta _{c}(4)=4$. The geometric description
together with the integral representation of $u_{x}(t,x)$ gives the
distribution $d\mu (t,\lambda )$ of the Lee--Yang zeroes at the scale $t$. $%
\Omega _{\infty }$ is a nonempty set and a nontrivial limit distribution is
attained but its support $\Sigma (t)$ is pushed away from the origin to
infinity.
\end{proposition}

\section{Normal Fluctuations\label{NF}}

\setcounter{equation}{0} \setcounter{theorem}{0}

We turn our attention to normal fluctuations. The block variable (\ref{sum})
is now normalized with $\gamma =d$ and the system is above the critical
temperature. The \textquotedblleft a priori\textquotedblright\ measure $%
\sigma _{K}^{(N)}(x)$, $L^{dK}=n$, that governs the law of (\ref{sum}),
satisfies a recursive equation 
\begin{equation*}
\sigma _{k}^{(N)}(x)=\frac{1}{C_{k}}e^{L^{-2k}(L^{\gamma }-1)\left\vert
x\right\vert ^{2}/2\gamma }\underset{L^{d}-\mathrm{times}}{\underbrace{%
\sigma _{k-1}^{(N)}\ast \cdots \ast \sigma _{k-1}^{(N)}}}(L^{d/2}x)\ ,\qquad
k\geq 1
\end{equation*}%
which, in view of $\gamma =d$, has an explicitly $k$ dependence in the
exponential pre--factor (see (\ref{sigmak})).

\medskip

\noindent \textbf{Initial value problem.} Following the procedure described
in Section \ref{I}, the initial value problem (\ref{V}) and (\ref{V0}), for
the logarithmic of its characteristic function $\phi _{k}^{(N)}(z)$ in the $%
L\downarrow 1$ limit, thus reads 
\begin{equation}
U_{t}=-\frac{1}{2}e^{-2t}\left( \Delta U-\left\vert U_{z}\right\vert
^{2}\right) +dU-\frac{\gamma }{2}z\cdot U_{z}+\frac{1}{2}e^{-2t}\Delta
U(t,0)\,.  \label{Ut}
\end{equation}%
Note that, $L^{-2k}=\exp \left( -2k\ln L\right) \rightarrow \exp \left(
-2t\right) $, as $k\rightarrow \infty $ together with $L\downarrow 1$ with $%
k\ln L=t$ fixed, and such function appears in front of the Laplacean in (\ref%
{rr}).

As $N\rightarrow \infty $, the radially symmetric solution of (\ref{Ut})
scaled properly satisfies the modified initial value problem (see (\ref{uU})
for the definition of $u(t,x)$):%
\begin{equation}
u_{t}=e^{-2t}u_{x}-2xe^{-2t}u_{x}^{2}-\gamma xu_{x}+du-e^{-2t}u_{x}(t,0)
\label{uu}
\end{equation}%
with $u(0,x)=u_{0}(x)$ given by (\ref{u0}).

We continue through equations (\ref{wu})-(\ref{up}). A similar Legendre
transform applied to (\ref{uu}) leads to the initial value problem 
\begin{equation}
v_{t}-2p^{2}e^{-2t}v_{p}=-e^{-2t}+\left( d+4e^{-2t}p\right) v  \label{ve}
\end{equation}%
with $v(0,p)=v_{0}(p)$ as given by (\ref{v0}). Note the cancellation of
terms proportional to $pv_{p}$ because $\gamma =d$ in this case.

\medskip

\noindent \textbf{Main result.} The following result holds for any $d>2$ by
it has been stated for $d=4$, for simplicity.

\begin{theorem}
\label{normal}Equations (\ref{ve}) and (\ref{v0}) with $d=4$ are solved by%
\begin{equation}
v(t,p)=\frac{-e^{4t}}{p^{2}}\left( 1-\frac{\beta }{4}+\frac{1}{p}\ln \left(
1-p+pe^{-2t}\right) +e^{-2t}\left( \ln \left( 1-p+pe^{-2t}\right) -1\right) -%
\frac{p}{2}e^{-4t}\right) ~.  \label{vetp}
\end{equation}%
For every $\beta <\beta _{c}(4)=4$ and $t\geq 0$, there is a unique solution 
$\bar{p}=\bar{p}(t,x)$ of%
\begin{equation}
v(t,p)=x~,  \label{vtx}
\end{equation}%
holomorphic in a neighborhood of the origin, that converges exponentially
fast, as $t\rightarrow \infty $, to the solution of%
\begin{equation*}
1-\frac{\beta }{4}=\frac{-1}{p}\ln \left( 1-p\right)
\end{equation*}%
in every compact set of $\mathbb{C}$. This implies, together with the
corresponding equations (\ref{up}) and (\ref{uV}), convergence to a Gaussian
equilibrium solution of the equation (\ref{Ut}) without terms proportional
to $e^{-2t}$:%
\begin{equation}
\lim_{t\rightarrow \infty }\lim_{N\rightarrow \infty }\frac{1}{N}U\left( t,%
\sqrt{N}z\right) =\frac{-\left\vert z\right\vert ^{2}}{2\mu (\beta )}
\label{UU}
\end{equation}%
uniformly in compact subsets of $-\left\vert z\right\vert ^{2}\in \mathbb{C}$%
.
\end{theorem}

\noindent \textit{Proof.} As in the proof of Theorem \ref{gauss}, equation (%
\ref{ve}) will be solved along the characteristics $p(t;p_{0})$. We refer to
this proof for details. Writing $V(t)=v(t,p(t))$, we have%
\begin{eqnarray}
\dot{p} &=&-2e^{-2t}p^{2}  \notag \\
\dot{V} &=&-e^{-2t}+\left( d+4e^{-2t}p\right) V  \label{ode}
\end{eqnarray}%
with initial conditions (\ref{pt0}). Integrating the first of these
equations gives%
\begin{equation}
p(t)=\frac{p_{0}}{1+p_{0}-p_{0}e^{-2t}}~.  \label{pet}
\end{equation}

The homogeneous equation $\dot{V}=\left( d+4e^{-2t}p\right) V$ can be
integrated analogously as before%
\begin{equation*}
V(t)=V_{0}e^{dt}\left( 1+p_{0}-p_{0}e^{-2t}\right) ^{2}
\end{equation*}%
Using the variation of constants formula, the solution to the second
equation of (\ref{ode}) is given by%
\begin{equation}
V(t)=e^{dt}\left( 1+p_{0}-p_{0}e^{-2t}\right) ^{2}\left( V_{0}-J_{0}\right)
\label{Vet}
\end{equation}%
with%
\begin{equation*}
J_{0}=\frac{1}{2}\int_{\exp \left( -2t\right) }^{1}\frac{\zeta ^{d/2}d\zeta 
}{\left( 1+p_{0}-p_{0}\zeta \right) ^{2}}.
\end{equation*}%
At this point, notice that $J_{0}$ for $d=4$ is exactly as in the proof of
Theorem \ref{gauss}. Equations (\ref{pt0}), (\ref{v0}) together with the
integration of $J_{0}$ gives 
\begin{equation*}
V_{0}+J_{0}=\left( \ref{V0J0}\right) ~.
\end{equation*}%
The difference between the two cases is the exponential pre--factor $e^{dt}$
of (\ref{Vet}) and $p_{0}=p_{0}(t,p)$ which is now obtained by solving (\ref%
{pet}) for $p_{0}$: 
\begin{equation}
p_{0}(t,p)=\frac{p}{1-p+pe^{-2t}}~.  \label{p0}
\end{equation}%
As we shall see, these two differences are responsible for the converge of
trajectories to different stationary solutions.

Equation (\ref{vetp}) follows by plugging (\ref{V0J0}) into (\ref{Vet}) with 
$p_{0}$ given by (\ref{p0}).

We now solve equation (\ref{vx}) for $p$ at $\beta \neq \beta _{c}=4$ which,
by (\ref{vtp}), can be written as 
\begin{equation}
\left( xp^{2}-\frac{p}{2}\right) e^{-4t}-e^{-2t}=-1+\frac{\beta }{4}-\left(
e^{-2t}-\frac{1}{p}\right) \ln \left( 1-p+pe^{-2t}\right) \equiv g_{1}(t,p)~.
\label{ge}
\end{equation}

Analogously to Lemma \ref{monotone}, we have

\begin{lemma}
\label{mono}For any $p<\left( 1-e^{-2t}\right) ^{-1}$, $g_{1}$ is a monotone
increasing function of $p$ with $g_{1}(t,0)=-e^{-2t}+\beta /4$ and diverges
logarithmically to $-\infty $ as $p\rightarrow -\infty $.
\end{lemma}

\noindent \textit{Proof of lemma.} $g_{1}$ is a monotone increasing function
of $p$ since%
\begin{equation*}
\left( g_{1}\right) _{p}(t,p)=\frac{e^{-2t}\left( 1-e^{-2t}\right) }{%
1-p\left( 1-e^{-2t}\right) }+\frac{1}{p^{2}}f\left( p\left( 1-e^{-2t}\right)
\right)
\end{equation*}%
with $f$ given by (\ref{fh}) is a positive function for .$p<\left(
1-e^{-2t}\right) ^{-1}$. Other statements follows as in the proof of Lemma %
\ref{monotone}.

\hfill $\Box $

The quadratic polynomial $Q_{1}(x,p)$ in the left hand side of (\ref{ge})
tends to a linear function $Q_{1}(0,p)=-e^{-2t}(1+pe^{-2t}/2)$ as $%
x\rightarrow 0$ with $Q_{1}(0,0)=-e^{-2t}$ and $Q_{1}(0,-2e^{2t})=0$. From
Lemma \ref{mono}, the graph of $g_{1}$ always intercepts the graph of $%
Q_{1}(0,p)$ for all $\beta >0$ and, as in the proof of Theorem \ref{gauss},
this implies the existence of a unique solution $\bar{p}(t,x)$ of (\ref{vtx}%
) for every $t\geq 0$, holomorphic in a neighborhood $U(t)$ of the origin
that becomes the entire complex plane $U(t)\rightarrow \mathbb{C}$ as $%
t\rightarrow \infty $. Details of the proof will be omitted since are
similar to the corresponding statements in Theorem \ref{gauss}.

\medskip

\noindent \textbf{Asymptotic expansion.} The asymptotic behavior of $\bar{p}%
\left( t,x\right) $ as $t\rightarrow \infty $ is given as follows. By
equation (\ref{ge}), $\bar{p}\left( t,x\right) $ converges exponentially
fast 
\begin{equation*}
\bar{p}\left( t,x\right) =\hat{p}\left( 1+\frac{4(\hat{p}+2)}{\beta -4\hat{p}%
(4-\beta )}e^{-2t}+O\left( e^{-4t}\right) \right)
\end{equation*}%
to a constant value $\hat{p}$ which solves 
\begin{equation}
1-\frac{\beta }{4}=\frac{-1}{\hat{p}}\ln \left( 1-\hat{p}\right) \equiv
h_{1}(\hat{p})~.  \label{beta1}
\end{equation}%
Since $h_{1}$ is a monotone increasing function of $\hat{p}<1$ with $%
h_{1}(0)=1$ and $\lim_{\hat{p}\rightarrow \infty }h_{1}(\hat{p})=0$, there
is a unique solution for all $0\leq \beta <4$. Comparing (\ref{beta1}) with (%
\ref{beta-mu}), together with (\ref{U}), (\ref{uV}) and (\ref{up}), equation
(\ref{UU}) holds with%
\begin{equation*}
\hat{p}=\frac{1}{2\mu (\beta )}~
\end{equation*}%
concluding the proof of Theorem \ref{normal}.

\hfill $\Box $

\section{Conclusions and Final Remarks\label{CR}}

\setcounter{equation}{0} \setcounter{theorem}{0}

In the present work, a continuous version of the hierarchical spherical
model at dimension $d=4$ has been investigated. The two main results are
Theorems \ref{gauss} and \ref{normal} on the limit distribution of the block
spin variable $X^{\gamma }$ normalized with exponent $\gamma =d+2$ at the
criticality and $\gamma =d$ above the critical temperature. To prove these
results, certain evolution equations corresponding to the renormalization
group transformation (\ref{rr}) in the limit $L\downarrow 1$ are solved
explicitly at $N=\infty $. Starting far away from the stationary Gaussian
fixed point the trajectories of these dynamical system pass through two
different regimes with distinguishable crossover behavior. The large--$N$
limit of the transformation (\ref{rr}) with $L^{d}$ fixed equal to $2$, at
the criticality, has been investigated in both weak and strong (coupling)
regimes by Watanabe \cite{W}. We mention that our analysis using the $%
L\downarrow 1$ limit equation is considerably simpler and, consequently, has
more details than Proposition $2.2$ in \cite{W}.

Theorem \ref{conformal} gives an interpretation for the above mentioned
trajectories using the geometric function theory. The methods used enable us
to describe the dynamics of the Lee--Yang zeroes along those trajectories.
As $N\rightarrow \infty $, the Lee--Yang zeroes becomes dense over a
semi--line and their measure, which depends on the scale parameter $t$, is
shown to reach a limit for $t$ large but the support of the limit measure is
pushed away to infinity as the trajectories approach the Gaussian fixed
point. The method also allow us to give the precise crossover scale $t_{%
\mathrm{co}}>t^{\ast }$ from strong to weak regime defined as the value of $%
t $ such that $\zeta _{t}^{\ast }=-\alpha (t)$ where $\zeta _{t}^{\ast }$ is
a fixed point of the function (\ref{vtp}) that solves equation (\ref{v}) and 
$-\alpha (t)$ is a point of the boundary of image domain $\Omega _{t}$.

There are, however, two major drawbacks in the $L\downarrow 1$ limit
equation of the hierarchical $O(N)$ Heisenberg model with $N$ finite.
Firstly, reflection positivity cannot be used to prove uniform convergence
of the $O(N)$ trajectories to $O(\infty )$ trajectories.

The other problem is related with the Lee--Yang property. A Borel\ measure $%
\rho $ in $\mathbb{R}^{N}$ possesses Lee--Yang property if its
characteristic function $\phi (z)=\displaystyle\int d\rho \left( x\right)
~\exp \left( iz\cdot x\right) $ belongs to the Laguerre class $\mathcal{L}$
of entire function of \ $\zeta =-\left\vert z\right\vert ^{2}\in \mathbb{C}$
which can be represented by%
\begin{equation}
f\left( \zeta \right) =\exp \left( \lambda \zeta \right) \prod_{k=1}^{\infty
}\left( 1+\frac{\zeta }{\alpha _{k}^{2}}\right)  \label{fzeta}
\end{equation}%
with $\lambda \geq 0$ and $\alpha _{1},~\alpha _{2},~\ldots $ real numbers
satisfying $\displaystyle\sum_{k=1}^{\infty }\alpha _{k}^{-2}<\infty $.
Hence (see \cite{N,HHW,W})%
\begin{equation}
h(\zeta )=-\zeta \left( \ln f\right) ^{\prime }(N\zeta )=\sum_{j=1}^{\infty
}\left( -1\right) ^{j}\nu _{2j}~\zeta ^{j}  \label{phiphi}
\end{equation}%
is holomorphic function of $\zeta $ in a neighborhood of the origin and
Newman's inequalities 
\begin{equation}
0\leq \nu _{2j}\leq \left( \nu _{4}\right) ^{j/2}  \label{nunu}
\end{equation}%
holds for all $j\geq 2$. The scaling (\ref{phiphi}) is chosen so that $\nu
_{2j}=O\left( 1\right) $ in $N$ for $j\geq 1$ if $\rho $ is the uniform
measure on the sphere of radius $\sqrt{N}$. Inequalities (\ref{nunu}) can be
shown to hold in the limit $N\rightarrow \infty $ but in this case $f$
cannot be represented by (\ref{fzeta}) as the zeroes $\left( \alpha
_{j}\right) _{j\geq 1}$ become dense over the real line.

Now, let 
\begin{equation*}
f_{k}=Tf_{k-1},k=1,2,\ldots 
\end{equation*}%
where $T:\mathcal{E}\longrightarrow \mathcal{E}$ is the operator defined by
recursion relation (\ref{rr}) with $f(\zeta )=\varphi \left( \left\vert
z\right\vert \right) =\phi (z)$, $\zeta =-\left\vert z\right\vert ^{2}$, be
a sequence in the space of entire functions $\mathcal{E}$ starting from $%
f_{0}(\beta \zeta )$ with $f_{0}$ in the Laguerre's class $\mathcal{L}$. It
has been proven in Theorem $1.1$ of \cite{KW} that, for every $k\in \mathbb{N%
}$ and $0\leq \beta \leq (L^{\gamma -d}-1)/\lambda $, 
\begin{equation*}
f_{k}\in \mathcal{L}\cap \mathcal{A}_{\lambda }
\end{equation*}%
where $\mathcal{A}_{a}$ denotes the Fr\'{e}chet space of functions $f\in 
\mathcal{E}$ such that 
\begin{equation*}
\left\Vert f\right\Vert _{b}:=\sup_{k\in \mathbb{N}}\frac{1}{b^{k}}%
\left\vert \frac{d^{k}f}{d\zeta ^{k}}(0)\right\vert 
\end{equation*}%
is finite for all $b>a$ and $\lambda $ is the type of $f_{0}$. This together
with equation (\ref{nunu}) can be used to establish the existence of a
critical inverse temperature $\beta _{c}$ such that the sequence $\left\{
f_{n}\right\} _{n\in \mathbb{N}}$ converges to $\exp \left( \zeta \right) $
uniformly in compact subsets of $\mathbb{C}$. The Pick class of functions is
the natural candidate for replacing Laguerre's class in the local potential
approximation of (\ref{rr}) but we don't have a substitute for the convex
space $\mathcal{A}_{\lambda }$. The present work is an attempt in this
direction for $N=\infty $.

\end{document}